\documentclass[lettersize,journal]{IEEEtran}
\usepackage{setspace}

\usepackage[dvipsnames]{xcolor}
\usepackage[normalem]{ulem} 

\usepackage{comment}
\usepackage[utf8]{inputenc} 
\usepackage[T1]{fontenc}
\usepackage[cmex10]{amsmath} 
\usepackage{subfigure}

\makeatletter\def\@captype{figure}\makeatother 

\makeatletter 
  \newcommand\figcaption{\def\@captype{figure}\caption} 
  \newcommand\tabcaption{\def\@captype{table}\caption}

\makeatother
\usepackage{amssymb}
\usepackage{multirow}
\usepackage{graphicx}

\usepackage{float}
\usepackage{comment}
\usepackage{color}
\usepackage{bm}
\usepackage[noadjust]{cite}
\usepackage{multirow}
\usepackage{enumerate}
\usepackage{epstopdf}
\usepackage{color}

\usepackage{ulem}

\usepackage{algorithm}
\usepackage{algorithmic}

\newcommand{\specialcell}[2][c]{%
  \begin{tabular}[#1]{@{}c@{}}#2\end{tabular}}
\allowdisplaybreaks
\ifCLASSINFOpdf
\else
\fi
\newtheorem{exmp}{Example}
\newtheorem{theorem}{Theorem}
\newtheorem{lemma}[theorem]{Lemma}

\newtheorem{remark}{Remark}

\newtheorem{definition}{Definition}

\begin{document}

\title{Analysis Methodology for Age of Information under Sequence Based Scheduling}
\author{{Fang Liu, Wing Shing Wong,~\IEEEmembership{IEEE Fellow}, Yuan-Hsun Lo,~\IEEEmembership{IEEE~Member}, \\ Yijin Zhang,~\IEEEmembership{IEEE~Senior Member}, Chung Shue Chen,~\IEEEmembership{IEEE Senior Member}}
\thanks{F. Liu is with College of Electronics and Information Engineering, Shenzhen University, Shenzhen.  Email: liuf@szu.edu.cn}
\thanks{W. S. Wong is with Department of Information Engineering, The Chinese University of Hong Kong (CUHK), Shatin, Hong Kong SAR. Email: wswong@ie.cuhk.edu.hk}
\thanks{Y.-H. Lo is with Department of Applied Mathematics, National Pingtung University, Taiwan.  Email: yhlo0830@gmail.com}
\thanks{Y. Zhang is with School of Electronic and Optical Engineering, Nanjing University of Science and Technology, Nanjing. Email: yijin.zhang@gmail.com}
\thanks{C. S. Chen is with Nokia Bell Labs,  Paris-Saclay Center, Massy 91300, France. Email: chung\_shue.chen@nokia-bell-labs.com}
\thanks{(\textit{Corresponding author: Wing Shing Wong})} }



\maketitle

\begin{abstract}
We focus on the Age of Information (AoI) performance in a system where each user generates packets periodically to send  to a common access point (AP) for status updating. To avoid heavy overhead, we assume that channel sensing, feedback information from the AP, and time synchronization are not available in the system. We adopt a multi-access scheme called the sequence scheme, where each user is assigned a periodic binary sequence to schedule their transmissions. 
In our previous work \cite{liu2023Age}, we have thoroughly studied the AoI performance under sequence scheme when the period of schedule sequences, $L$, is equal to the status generating period, $T$. The results can be extended to the case where $T>L$. However, the case of $T<L$ is not covered by \cite{liu2023Age}. Therefore, in this paper, 
we concentrate on analyzing the AoI performance in the case of  $T<L$, which is more challenging and requires different approaches. 
We conduct in-depth analysis on this case and develop a mathematical tool based on integer partitions to facilitate the analysis. We derive low-complexity closed-form expressions for two scenarios under $T<L$. 
Based on the obtained analytical results, we propose an algorithm to optimize the construction parameters of the sequence scheme. Finally, we compare our proposed sequence scheme with two commonly used baselines, and show that our proposed scheme outperforms the baselines in terms of AoI performance while consuming less energy.
\end{abstract}


\section{Introduction}
\IEEEPARstart{T}{he} proliferation of IoT devices has significantly boosted the demand for time-critical information updates in a broad range of applications, such as real-time traffic data exchange for autonomous-driving vehicles and instantaneous user utility profiles for smart grid control. The Age of Information (AoI) metric was first proposed by \cite{kaul2011minimizing}  to   capture the information timeliness  in real-time systems.  It is defined as the time elapsed since the generation time of the latest received status update at the receiver side. 
Complementing conventional performance metrics such as throughput and latency, such a definition highlights the importance of keeping information available at the receiver side as fresh as possible.

Extensive research on AoI has been conducted under various network settings  \cite{ding2023age,asvadi2023peak,chen2022age}. In this paper, we consider an application scenario in which multiple users are required to send periodically generated status packets to a common access point (AP) by sharing a single channel. The generation period for status packets, referred to as the \textit{frame length}, is denoted by $T$. This model is frequently encountered in practical applications. For example, in industrial manufacturing, sensors deployed in factories periodically report the operating status of equipment to a control center to enhance production efficiency \cite{librino2020resource,peng2022delay}. Similar scenarios can be observed in periodic wireless sensor networks (PWSNs) \cite{bae2022age,zhang2022aoi}, vehicle-to-infrastructure (V2I) communications \cite{chen2021adaptive},  and other~domains.

The AoI performance of users is significantly affected by how multi-accessing is handled. Carrier sense multiple access (CSMA), in which users listen for carrier signals before transmitting their own data to avoid collisions, is a popular choice \cite{maatouk2020age,zhou2021performance,tripathi2023fresh}. 
However, carrier sensing may cause high overhead, especially for small packet transmission. As pointed out in \cite{woo2001transmission,nguyen2020distributed}, control packets in CSMA, such as RTS, CTS, and ACK, may contribute up to 40\% of the total overhead costs. Therefore, we are motivated to consider systems that do not employ carrier sensing.
There are also many multi-access schemes relying on the availability of feedback information from the AP \cite{yang2021game,bae2022age}.
Despite potential performance improvement, feedback transmission usually relies on a dedicated control channel, which can become a bottleneck in heavy traffic situations and is not suitable for energy-constrained networks such as sensor networks. Thus, we also assume that there is no such feedback from the AP in our model. Additionally, we assume that time synchronization among the users is not available in the system to further reduce overhead.

  
Although probabilistic multi-access schemes are still available under these constraints, in this paper, we mainly focus on deterministic schemes. We view deterministic schemes as being defined through sequences, wherein each user is pre-assigned a transmission schedule represented by a periodic binary sequence of length $L$. At each time slot, each user reads its current sequence value and takes an action accordingly, either transmitting through the channel or remaining silent. The sequence schemes are designed to operate without relying on channel monitoring or feedback information from the AP. This simplicity in implementation is particularly advantageous under our model assumptions. 
Moreover, as indicated in the literature, \cite{chen2018crt,shum2010construction}, there exist schedule sequence sets that can guarantee at least one collision-free transmission within a sequence period for all asynchronous users. In contrast, many other well-known methods fail to offer this guarantee due to their probabilistic nature.


Our earlier paper \cite{liu2023Age} provides the first analysis for AoI under sequence scheme without time synchronization. In that work,
we derived an explicit formula for  average AoI under the assumption that the status updating period and the sequence period are identical, that is, $T=L$. Additionally, we characterized critical sequence properties
that are essential for AoI optimization.
This paper extends our former results in two key aspects:
First, we remove the restriction on the sequence period. In our previous work, we mainly concentrated
on the $T=L$ case. Although extension to the $T > L$ case is relatively straightforward
as observed in \cite{liu2023Age}, the $T < L$ case, which is the main focus of this work, cannot be covered by \cite{liu2023Age}. 
We note that this case is more challenging to analyze and   requires approaches that are quite different from those in~\cite{liu2023Age}. 
This is because  when $T \geq L$, the transmission patterns within different frames are identical for a given sequence. However, when $T < L$, the transmission pattern varies from frame to frame, making the analysis more complex.
Second, the optimization analysis in \cite{liu2023Age} focused on selecting
the optimal cyclical shift parameter after sequence construction. In this work, we extend our
optimization consideration to include parameter selection before sequence construction. 


The significance of studying the $T < L$ case can be attributed to two primary factors: 1) Since sequence length $L$ is typically quadratic with respect to the number of users, in systems with frequent status updates, it is more likely that the status generation period $T$ falls within the range $[1, L-1]$ rather than $[L, \infty)$. This makes the $T < L$ case more common in practical applications. 2) By studying the $T < L$ scenario,  we fill a gap in the literature and provide a holistic view  of the AoI performance under sequence schemes for general values of $T$. Our study offers valuable insights and guidance for system design and optimization across various configurations.

The main contributions of this paper are summarized below.
\begin{enumerate}
\item 
We conduct analysis on AoI under the sequence scheme for the case where $T<L$, in the context of time asynchronization. This covers a different parameter range compared to our previous work in \cite{liu2023Age} which is on $T \geq L$. 
Using direct definition (shown in \eqref{eq:AOI}) to compute the average AoI for arbitrary starting time offsets causes prohibitively high computation complexity.
To address this challenge, we derive two new formulas (shown in \eqref{eq:A in S and Y},  \eqref{eq:A in S and X}) to  improve the 
computational efficiency. 
Besides, we
develop a new mathematical approach based on 
integer partitions, to evaluate complex statistical expressions that arise in the AoI calculation.

\item

We conduct an in-depth study for two specific cases: 1)   $T$ and $L$ are co-prime, and 2) each frame contains at most one transmission slot. For these  cases, our proposed  approach leads to closed-form expressions  with significantly lower complexity, $O(N^4)$. 
The formulas are consistent  with the simulation results.

\item 
Our analytical results  provide valuable insights in understanding the AoI performance of the system.
Based on these results, we propose an optimization method for parameter selection in sequence construction to minimize the average AoI.

\item 
We compare the sequence scheme with two commonly used baselines, slotted ALOHA and framed ALOHA, by numerical study. 
Results show that the sequence scheme achieves enhanced AoI performance. 
Moreover, we evaluate the energy consumption of the schemes and find that the sequence scheme outperforms the baselines in terms of energy~efficiency. 


\end{enumerate}

The rest of this paper is organized as follows. Related works on AoI study are introduced in Section \ref{sec:related work}. 
After describing the system model in Section~\ref{sec:system model},  we present preliminary technical details for sequence design in Section~\ref{sec:CRT sequence} to prepare for 
subsequent discussions. We conduct analysis  on average AoI  and develop mathematical tools in Section~\ref{sec:general results}. 
Low-complexity closed-form expressions for two special cases are derived in Section~\ref{sec:specail cases}. Based on the analytical results, we propose AoI optimization algorithm in Section~\ref{sec:AoI optimization}.
The simulation results are shown in   Section~\ref{sec:simulation}. Finally, we conclude the paper in Section~\ref{sec:conclusion}.

\section{Related Works}\label{sec:related work}
AoI has garnered significant research interest  since its introduction in the context of vehicular ad hoc networks (VANETs)
by \cite{kaul2011minimizing}. 
Here we review AoI-related studies in multi-access scenarios, categorizing them based on their approach: random or deterministic.

Various random schemes have been proposed to enhance AoI performance in multi-access scenarios, many of which are based on CSMA. In CSMA, users conduct carrier sensing before transmitting their own data to avoid collisions.
In \cite{maatouk2020age}, the authors investigate AoI under CSMA and analyze the optimal backoff times for minimizing AoI when updates are generated at will. In \cite{zhou2021performance}, a mean-field
approximation approach with guaranteed accuracy is developed to analyze the asymptotic AoI performance for different CSMA-based schemes under Poisson arrival traffic. A distributed CSMA protocol is proposed in \cite{tripathi2023fresh} to minimize an AoI weighted sum by replicating the behavior of centralized policies that are known to be nearly optimal.
Many random schemes depend on feedback information from the AP. For example, authors in \cite{chen2019age} assume that a receiver immediately broadcasts an error-free acknowledgment for all successful packets that are received in a time slot. In \cite{yang2021game}, probabilistic feedback due to varying channel environment is studied. In the slotted ALOHA scheme presented in \cite{bae2022age}, the number of contending users is announced by the AP to enable the users to adjust the transmitting probability in each slot accordingly. There are a variety of learning-based schemes in which feedback information on AoI is crucial for the state representation and decision-making process in learning algorithms \cite{zakeri2023minimizing,liu2021average,peng2023aoi}.

For AoI studies under deterministic multi-access schemes, most of them are conducted under the assumption of time synchronization.  For instance, \cite{li2023eywa} introduces the design of  sequence schemes with two objectives: minimizing weighted AoI and reducing bandwidth under AoI constraints. Additionally, \cite{li2022scheduling} proposes a method for constructing low-complexity sequences based on a specified maximum AoI threshold. 
Our earlier work \cite{liu2023Age} offers the first study for AoI under sequence scheme in asynchronous scenario.
In that work, our derivation for closed-form formula on average AoI and proposed AoI optimization algorithm are developed based on the assumption that the status updating period and the sequence period are identical, that is, $T=L$. Compared to  \cite{liu2023Age}, this paper concentrates on addressing the case where 
 $T>L$.

We focus on the same sequences as in   \cite{liu2023Age}, that is, MHUI sequences.
They are a type of protocol sequences initially proposed for transmission scheduling on collision channels without feedback. MHUI sequences offer a much shorter sequence length compared to other protocol sequences, such as shift-invariant (SI) sequences \cite{shum2009shift} and wobbling sequences \cite{wong2007new}, while still providing a minimum throughput guarantee, that is, each sequence in an MHUI set is guaranteed to contain at least one collision-free ``1'' within a sequence period.
MHUI sequence sets with a uniform Hamming weight are also recognized as conflict-avoiding codes (CAC)~\cite{jimbo2007conflict,hsia2024conflict}, which have been extensively studied in the literature. Additionally, MHUI sequence sets are also closely related to another combinatorial structure called cover-free family~\cite{erdos1985families,alon2020probabilistic}. 
MHUI sequences have been explored in various application scenarios. For instance, \cite{wu2014safety} employs MHUI sequences for safety message broadcast in the scenario of VANETs, and \cite{chen2018crt} considers the applications of MHUI sequences to collision channels allowing successive interference cancellation (SIC) at the receiver.

\section{System Model} \label{sec:system model}

\subsection{Overall Scenario}

This paper investigates a  time-slotted multi-access system consisting of $N$ users and a common AP. The system operates in discrete time slots, indexed by $t=0,1,2,\ldots$.  
At the beginning of each frame, each user generates a packet carrying status information to be reported to the AP. The frame length is denoted by~$T$. Each packet should be delivered to the AP successfully 
within the frame it is generated, otherwise it will expire and be discarded at the end of the frame. All packets from users to the AP are transmitted over a perfect collision channel that is mutually shared. If only one user transmits through the channel at a given time slot, then the transmitted packet is successfully received by the AP. However, if more than one user transmits in the same slot, then a collision occurs and no packets
sent at this time slot can be successfully received. Furthermore, we assume that feedback information from AP regarding whether a transmission is successful or not is not available in the system. The transmission delay of each packet from a user to the AP is assumed to be bounded by the length of a slot. 

The users start their transmission sessions independently and do not share a common  time clock. 
It follows that their local clocks may differ from each other.
Without loss of generality, we assume that there exists a system reference time. 
For $i\in \{1,2,\ldots,N \}$, we denote the starting time offset between the reference time and user~$i$'s local time by $\tau_i$, whose value is unknown beforehand. 
A collection of $\tau_i$'s is denoted by an \emph{offset vector} $\bm{\tau}=(\tau_1, \tau_2, \ldots, \tau_N) $. 
 To facilitate discussions, we assume that the slot boundaries of different users are aligned, and then the starting time offsets can only take values from non-negative numbers. 
 This assumption does not limit the generality of our study, as previous studies have demonstrated that the results can be easily extended to the non-aligned case~\cite{chang2019asynchronous,liu2023Age}. 


\subsection{Sequence Based Scheduling}
\begin{figure}[t] 
 \begin{minipage}[b]{0.5\textwidth} 
    \centering
\includegraphics[width=1\textwidth]{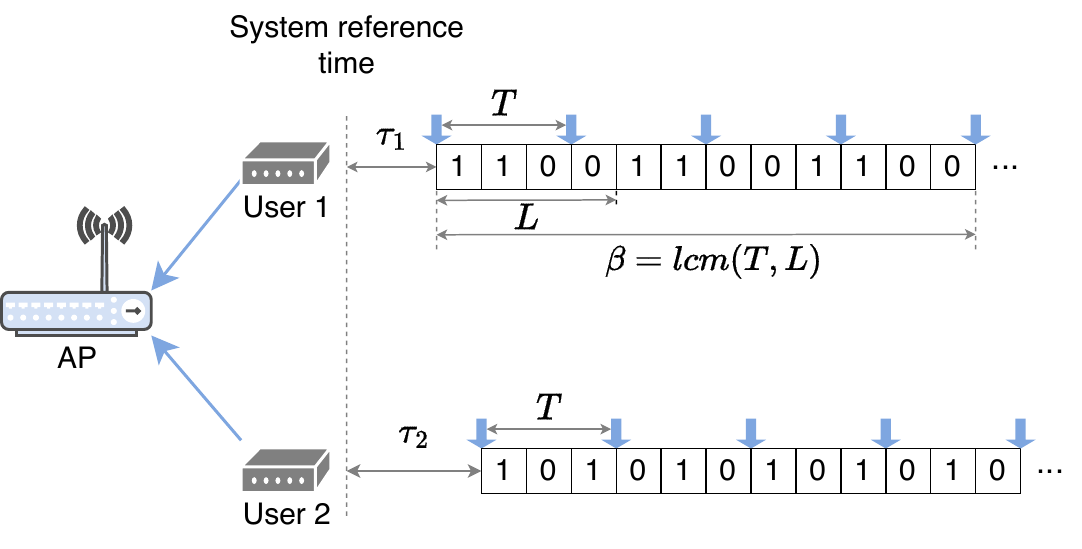} 
	\caption{Illustration for the sequence scheme. The two users  are assigned with sequences $\bm{s}_1=[1\ 1\ 0\ 0]$ and $\bm{s}_2=[1\ 0\ 1\ 0]$, respectively. The sequence period equals $L=4$, and the frame length equals $T=3$. It follows that the length of a superframe equals $\beta=12$. } 
	\label{Fig.System} 
  \end{minipage}%
  \hspace{0.2cm}
   \begin{minipage}[b]{0.5\textwidth} 
    \centering
       \includegraphics[width=0.88\textwidth]{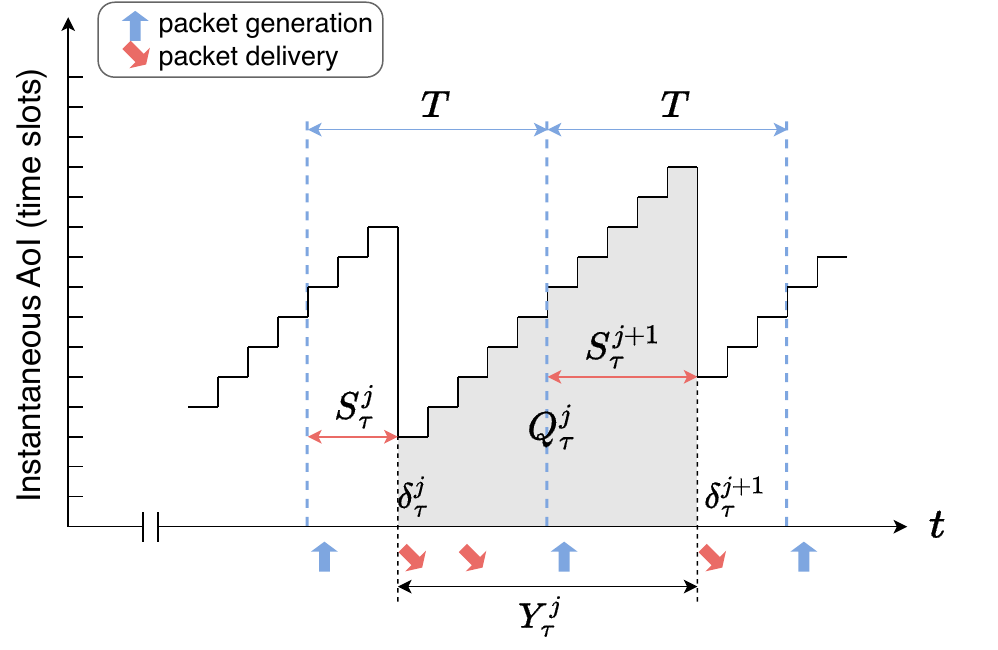} 
	\caption{ Illustration for the time evolution of $A_{\bm{\tau}}(t)$. As observed,  $A_{\bm{\tau}}(t)$ grows linearly over time in the absence of successful update delivery. Conversely,  when a packet sent by user $i$ is delivered successfully, then $A_{\bm{\tau}}(t)$ immediately drops to the service time of this packet.} 
	\label{Fig.AoI}    
  \end{minipage}%
\end{figure}

We consider a sequence based multi-access scheme, in which each user is pre-assigned a unique  binary and periodic sequence to schedule its packet transmissions. 
The sequences are denoted by $\{\bm{s}_1,\bm{s}_2,\ldots,\bm{s}_N\}$, and the period of these sequences is  denoted by $L$.
The sequence assigned to user $i$ is denoted by
$\bm{s}_i= [s_i (0) \ s_i(1) \ \ldots \ s_i(L-1)], $
where $ s_i(j)\in \{0,1 \}$, $j\in \mathbb{Z}_L$, for $i\in \{1,2,\ldots, N\}$. 
At each time slot, each user reads out its sequence value, and then transmits a packet
when the sequence value is 1 and  keeps silent otherwise. Due to the periodicity of the sequences, we assume $\tau_i \in \mathbb{Z}_L$ without loss of generality.
Let $\beta=\operatorname{lcm}(T,L)$ be the least common multiple of $T$ and $L$. 
We define a \emph{superframe} of length~$\beta$ as a  concatenation of $\beta/T$ frames. 
The sequence of user~$i$ within a superframe is denoted by $\bm{r}_i=\underbrace{[\bm{s}_i\ \ldots \ \bm{s}_i]}_{\beta/L}$. Fig.~\ref{Fig.System}  illustrates a superframe under the sequence~scheme. 

We focus on  a common class of schedule  sequences,  namely Minimum Hamming User-Irrepressible (MHUI) sequences, as in \cite{liu2023Age}. The MHUI sequence set enjoys the property that each sequence in the set has at least one collision-free ``1'' within a period $L$, no matter what the starting time offsets are. The characteristics of MHUI sequences will be introduced in detail in Section~\ref{sec:CRT sequence}. 

\subsection{Age of Information }
\label{subsec:AoI}
The AoI of a user measures the timeliness of the status updates at the AP side, representing the time elapsed since the generation of the last successfully received status. 
Since all users have identical traffic pattern and use the same multi-access protocol,  we can focus our analysis on a single user, user~$i$, $i\in \{1,2,\ldots,N \}$, without loss of generality. 
We may omit the user index, $i$, in some  notations for brevity.
The \emph{instantaneous AoI} of user $i$ at time slot $t$ under a given offset vector $\bm{\tau}$ is denoted by $A_{\bm{\tau}}(t)$. 
We illustrate the time evolution of $A_{\bm{\tau}}(t)$ in Fig.~\ref{Fig.AoI}.
As depicted in Fig.~\ref{Fig.AoI}, when there is no successful update delivery from user $i$, $A_{\bm{\tau}}(t)$ increases linearly over time according to $A_{\bm{\tau}}(t) = A_{\bm{\tau}}(t-1) + 1$. 
Conversely, when a packet sent by user $i$ is delivered successfully, the AP updates user $i$'s state accordingly, resulting in a \emph{drop} in AoI to $A_{\bm{\tau}}(t) = t - g_{\bm{\tau}}(t)$, where $g_{\bm{\tau}}(t)$ represents the time at which this packet was generated.

Note that in the system, a packet may be transmitted successfully more than once before its expiration. 
It should be pointed out that only the first successful transmission of a status packet would cause a drop of instantaneous AoI, 
while subsequent successful transmissions of the same packet, if any, would not. 
 For user $i$, we denote the time slot when its instantaneous AoI experiences the $j$-th drop by $\delta_{\bm{\tau}}^j$, $j=1,2,\ldots$.
For the packet successfully delivered at $\delta_{\bm{\tau}}^j$, we define its \emph{service time} $S_{\bm{\tau}}^j$  as the time interval between its generation time  and $\delta_{\bm{\tau}}^j$. 
Besides, we use $Y_{\bm{\tau}}^j$ to denote the \emph{inter-departure time} between two consecutive AoI drops, i.e., $Y_{\bm{\tau}}^j=\delta_{\bm{\tau}}^{j+1}-\delta_{\bm{\tau}}^j, j=1,2,\ldots $.

The \emph{average AoI} of user~$i$ for a given offset vector $\bm{\tau}$, denoted by $\overline{A_{\bm{\tau}}}$,  is defined as
$
   \overline{A_{\bm{\tau}}}=\lim \limits_{\lambda \rightarrow \infty}\frac{1}{\lambda} \sum_{t=0}^{\lambda}A_{\bm{\tau}}(t)$. 
Under the sequence scheme, the instantaneous AoI of user $i$ changes periodically with period~$\beta$. 
We restrict our attention  within a superframe of length~$\beta$, which consists of $\beta/T$ frames. 
That is, we have  
$
 \overline{A_{\bm{\tau}}}   = \frac{1}{\beta} \sum_{t=0}^{\beta-1}A_{\bm{\tau}}(t) $.
Let $\eta$ be the total number of AoI drops within a superframe, and $Q_{\bm{\tau}}^j$ be the sum of instantaneous AoI between $\delta_{\bm{\tau}}^j$ and $\delta_{\bm{\tau}}^{j+1}$, as exemplified in the gray polygon area of Fig.~\ref{Fig.AoI}. Then, we have
$
  \overline{A_{\bm{\tau}}}    = \frac{\eta}{\beta} \frac{1}{\eta} \sum_{t=0}^{\beta-1}A_{\bm{\tau}}(t)
=\frac{\mathbb{E}_j[Q_{\bm{\tau}}^j]}{\mathbb{E}_j[Y_{\bm{\tau}}^j]}.  $
The area $Q_{\bm{\tau}}^j$ can be calculated as
$
Q_{\bm{\tau}}^j=S_{\bm{\tau}}^j+(S_{\bm{\tau}}^j+1)+\cdots+(S_{\bm{\tau}}^j+Y_{\bm{\tau}}^j-1) 
	=S_{\bm{\tau}}^{j}Y_{\bm{\tau}}^j+\frac{({Y_{\bm{\tau}}^j})^2-Y_{\bm{\tau}}^j}{2}.
$
It follows that \cite{chen2022age} 
$
\overline{A_{\bm{\tau}}}=\frac{\mathbb{E}_j[Q_{\bm{\tau}}^j]}{\mathbb{E}_j[Y_{\bm{\tau}}^j]}=\frac{\mathbb{E}_j[S_{\bm{\tau}}^{j}Y_{\bm{\tau}}^j]}{\mathbb{E}_j[Y_{\bm{\tau}}^j]}+\frac{\mathbb{E}_j[({Y_{\bm{\tau}}^j})^2]}{2\mathbb{E}_j[Y_{\bm{\tau}}^j]}-\frac{1}{2}.$ 
The exact value of $\bm{\tau}$ is unknown due to the lack of time synchronization. The sequence scheme we propose is influenced by the variability in offsets, leading to significant differences in AoI performance. 
To account for this variability, we take the average of $\overline{A_{\bm{\tau}}}$ over all $\bm{\tau}$, denoted by $\overline{A}=\mathbb{E}_{\bm{\tau}} \left[ \overline{A_{\bm{\tau}}}\right]$, as shown below.
 \begin{equation}\label{eq:AOI}
 \overline{A}=\mathbb{E}_{\bm{\tau}} \left[ \overline{A_{\bm{\tau}}}\right]=\mathbb{E}_{\bm{\tau}} \left[ \frac{\mathbb{E}_j[S_{\bm{\tau}}^{j}Y_{\bm{\tau}}^j]}{\mathbb{E}_j[Y_{\bm{\tau}}^j]}+\frac{\mathbb{E}_j[({Y_{\bm{\tau}}^j})^2]}{2\mathbb{E}_j[Y_{\bm{\tau}}^j]}-\frac{1}{2}\right].
 \end{equation}
 This approach provides a robust assessment of the system’s performance across different offset scenarios.

In this paper, our focus is on deriving $\overline{A}$ under the MHUI sequence scheme with $T<L$. The most straightforward approach is to enumerate all possible cases of $\bm{\tau}$, which is uniformly distributed and has a space size of $L^N$. 
For each $\bm{\tau}$, we use $O(\rho)$ to denote the computational complexity of $\overline{A_{\bm{\tau}}}$. Consequently, this approach has a computational complexity of $O(L^N \rho)$, which can be prohibitively high especially when $N$ is large, as $L$ is usually positively correlated with $N$. 
Therefore, we aim at deriving $\overline{A}$ in a  way of low computational complexity.  
As we will show in later sections, we can reduce the computational complexity to $O(N^4)$ in certain~cases. 

\section{Preliminaries on MHUI Sequences} \label{sec:CRT sequence}
In this section, we introduce mathematical preliminaries to facilitate the analysis on AoI under MHUI sequences, including  essential definitions and critical properties of MHUI sequences. The construction method of MHUI sequences is also provided. 
The main notations used in this paper are shown in Table~\ref{tab:notation}. 

\begin{table}[t]
\center
\caption{Notation  Table}
\label{tab:notation}
\scalebox{0.75}{
\begin{tabular}{c|c}
\hline
Notation  & Definition \\
\hline
 $ N $ & Total number of users  \\
 \hline
 $T$ & Frame length \\
 \hline
 $ L $ & Period of a schedule sequence set \\
 \hline
  $ \beta $ &  Length of a superframe, $\beta= \operatorname{lcm} (T,L)$\\
  \hline
$\tau_i$ & Starting time offset  of user $i$, $\tau_i \in \mathbb{Z}_{L}$  \\
\hline
 $\bm{\tau}$ & Starting time offset vector, 
 $ \bm{\tau} =[\tau_1,\tau_2,\ldots, \tau_N] \in \mathbb{Z}_{L}^{N} $\\
 \hline
 $ \bm{s}_i $ & Schedule sequence assigned to user $i$   \\
 \hline
$\bm{r}_i$ &  Schedule sequence of user~$i$ within a superframe $\bm{r}_i=\underbrace{[\bm{s}_i\ \ldots \ \bm{s}_i]}_{\beta/L}$ \\
\hline
 $w$ & Hamming weight of $\bm{s}_i$ \\
 \hline
 $w'$ & Hamming weight of $\bm{r}_i$ \\
 \hline
 $\ell_k$ & \specialcell{Cyclic distance between the $k$-th  and $(k \oplus_{w'} 1)$-th ``1''s in $\bm{r}_i$, \\ $k\in \mathbb{Z}_{w'}$ }\\
 \hline 
 $\sigma_k$ & 1-position of the $k$-th ``1'' in $\bm{r}_i$, $k\in \mathbb{Z}_{w'}$\\
 \hline
  $\mathcal{D}$ & 1-position set of  $\bm{r}_i$, $\mathcal{D}=\{\sigma_0,\sigma_1,\ldots,\sigma_{w'-1}\}$\\
  \hline
$\mathcal{E}_r$ & \specialcell{Set of events in each of which the number of \\ successful ``1''s in $\bm{s}_i$ is equal to $r$ } \\
\hline
$d^e_k$ & \specialcell{Cyclic distance between the $k$-th and  $(k\oplus_r 1)$-th successful ``1''s \\ in $\bm{s}_i$  under event $e$,  $k\in \mathbb{Z}_r $, $e\in \mathcal{E}_r$ } \\
\hline
 $S^e_j$ & \specialcell{Service time of the packet delivered at the $j$-th AoI drop \\ under event $e$, $e\in \mathcal{E}_r$} \\ 
 \hline
 $Y^e_j$ & \specialcell{Slot-level inter-departure time between  the $j$-th  and  \\ $(j+1)$-th AoI drops  under event $e$, $e\in \mathcal{E}_r$ }\\ 
 \hline
  $X^e_j$ & \specialcell{Frame-level inter-departure time between  the $j$-th  and \\ $(j+1)$-th AoI drops  under event $e$, $e\in \mathcal{E}_r$ }\\ 
 \hline
 $\overline{A}$ & Average AoI (AAoI) of user $i$ over all possible $\bm{\tau}$s\\
\hline
\end{tabular}}
\end{table}


The \emph{Hamming weight} of a binary sequence is defined as the number of ``1''s within it.
To facilitate discussions, we consider MHUI sequences $\{\bm{s}_1,\bm{s}_2,\ldots,\bm{s}_N \}$ with identical Hamming weight, denoted by $w$. 
The Hamming weight of each $\bm{r}_i$ is denoted by $w'$, which is equal to $w \beta /L$.
We use the set $\mathcal{I}_i=\{x_0,x_1,\ldots,x_{w-1} \}$, called \emph{characteristic set}, to represent the positions of the $w$ ``1''s within $\bm{s}_i$, $x_b \in \mathbb{Z}_L$, $b \in \mathbb{Z}_w$. 
The characteristic set of~$\bm{r}_i$ is denoted by $\mathcal{I}'_i=\{y_0,y_1,\ldots,y_{w'-1} \}$. 
Based on $\mathcal{I}_i$, we have $\mathcal{I}'_i=\{aL+x_b: a \in \mathbb{Z}_{\beta/L}, b\in \mathbb{Z}_w  \}$. 
We use $\ell_k$ to denote the \emph{cyclic distance} between the $k$-th ``1'' and the $(k \oplus_{w'} 1)$-th ``1'' in $\bm{r}_i$, $k\in \mathbb{Z}_{w'}$. 
That is,  $\ell_{k}=y_{k+1}-y_k$ for $k=0,1,\ldots,w'-2$, and $\ell_{k}=y_0+\beta-y_{w'-1}$ for $k=w'-1$.

\begin{definition} \label{def:MHUI}
\textbf{MHUI sequence set} \cite{liu2023Age}:
A set of $N$ binary sequences of a common period, say $L$, is called a \emph{minimum Hamming user-irrepressible (MHUI)} sequence set if it satisfies: 1) the Hamming weight of each sequence is no less than $N$, and 2)  the Hamming cross-correlation between any two sequences in the set is no more than 1 for any starting time offsets, that is,
$
H_{i,j}(\tau) =\sum_{t=0}^{L-1}s_{i}(t)s_{j}(t\oplus_{L}\tau) \leq 1, 
$
for  $\bm{s}_i= [s_i (0) \ s_i(1) \ \ldots \ s_i(L-1)] $, $\bm{s}_j= [s_j (0) \ s_j(1) \ \ldots \ s_j(L-1)] $, $ i,j \in \{1,2,\ldots,N\}$, $i\neq j$,  and $\tau\in\mathbb{Z}_L$.
\end{definition}

According to Definition~\ref{def:MHUI}, each sequence in an MHUI sequence set is guaranteed to have at least one collision-free ``1'' within a sequence period, regardless of the starting time offsets. This characteristic is known as the \textit{User-Irrepressible (UI)} property. Sequence sets that exhibit this property are referred to as \textit{UI sequence sets} \cite{shum2009design,chen2018crt}. Therefore, MHUI sequence sets are a family of UI sequence sets.
In the following, we introduce a most common construction method for MHUI sequences, which is based on the Chinese Remainder Theorem~(CRT).

\begin{definition} \label{def:CRT-UI}
\textbf{CRT construction} \cite{shum2010construction}: 
Given $N$, let $p$ be the smallest prime satisfying $p\geq N$, $q$ be an integer coprime with $p$ satisfying $q \geq 2p-1$, and $w=p$. 
As $\gcd(p,q)=1$, there is a natural bijective mapping  $\Phi_{p,q}:\mathbb{Z}_{pq}\to\mathbb{Z}_{p} \times \mathbb{Z}_{q}$ by $\Phi_{p,q}(t)=(t \text{ mod }  p, t  \text{ mod } q)$, which is the well-known \emph{Chinese Remainder Theorem (CRT) correspondence}.
We construct a set of $(p+1)$ sequences $\{ \bm{v}_{g}=[v_g(0)~v_g(1)~\ldots ~v_g(L-1)]: g\in \{1,2,\ldots,p+1\} \}$ with a common period $ L=pq $ as follows. 
For $g\in \{1,2,\ldots,p\} $, $t\in \mathbb{Z}_L$, define
\begin{equation}  \label{eq:CRT-seq1}
v_{g}(t)=\begin{cases}
1, \text{if } \Phi_{p,q}(t)= (ug \ \text{mod}\ p, u\ \text{mod}\ q), \text{ for }u\in \mathbb{Z}_{w}, \\
0, \text{otherwise}.
\end{cases}
\end{equation}
When $g=p+1$, for $t\in\mathbb{Z}_L$, define
\begin{equation} \label{eq:CRT-seq2}
    v_{g}(t)=\begin{cases}
1, \text{if } \Phi_{p,q}(t)= (u \ \text{mod}\ p, 0), \text{ for }u\in \mathbb{Z}_{w}, \\
0, \text{otherwise}.
\end{cases}
\end{equation}
\end{definition} 

Any $N$ sequences out of $\{\bm{v}_1, \bm{v}_2, \ldots, \bm{v}_{p+1}\}$ form a set of $N$ MHUI sequences $\{\bm{s}_1, \bm{s}_2, \ldots, \bm{s}_{N}\}$~\cite{shum2009design}.

\begin{exmp} \label{example 1}
Given $ N=3 $, we design four sequences by the CRT construction with   $ p=3 $, $ q=5 $, $ w=3 $ and  $ L=pq=15 $ as follows. Any three of them form a set of three MHUI~sequences.
\begin{align} \label{example_s2}
&\bm{v}_{1}=[1 1 1 0 0 0 0 0 0 0 0 0 0 0 0], 
\bm{v}_{2}=[1 0 0 0 0 0 0 1 0 0 0 1 0 0 0],\\
&\bm{v}_{3}=[1 0 0 0 0 0 1 0 0 0 0 0 1 0 0],
\bm{v}_{4}=[1 0 0 0 0 1 0 0 0 0 1 0 0 0 0]. \notag
\end{align}

\end{exmp}

Recall that the characteristic set $\mathcal{I}'_i=\{y_0,y_1,\ldots,y_{w'-1} \}$ indicates the positions of the  $w'$ ``1''s  within $\bm{r}_i$. 
To study the AoI performance, it is essential to understand the position of each ``1'' within its individual frame. 
This raises the following definition.
\begin{definition} \label{def:1-position}
\textbf{1-position}. 
Consider the characteristic set $\mathcal{I}'_i=\{y_0,y_1,\ldots,y_{w'-1} \}$ of $\bm{r}_i$.
For the $k$-th ``1'' in $\bm{r}_i$, $k\in \mathbb{Z}_{w'} $, define its \emph{1-position}, denoted by $\sigma_k$, as the position of this ``1'' within its individual frame. That is, $\sigma_k \equiv y_k \ \text{mod} \ {T}$.
\end{definition}

\begin{figure}[t] 
	\centering 
	\includegraphics[width=0.48\textwidth]{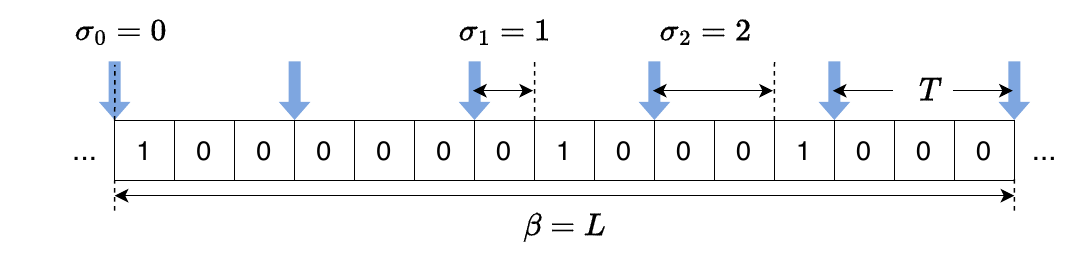} 
	\vspace{-0.2cm}
	\caption{Illustration of 1-positions. 
 In the given scenario, $L=15, T=3$,  $\beta=15$, and the sequence used is $  \bm{v}_2$ in \eqref{example_s2}, which is of Hamming weight $w=3$. Within a superframe, we have $w'=w=3$, and   $\mathcal{I}'_i=\mathcal{I}_i=\{0,7,11 \}$. The $w'$ ``1''s  correspond  to $\sigma_0=0, \sigma_1=1, \sigma_2=2$, respectively. Thus, $\mathcal{D}=\{0,1,2 \}$.
 } 
	\label{Fig.1-different} 
		\vspace{-0.2cm}
\end{figure}

Define the set of 1-positions within a superframe $\bm{r}_i$ by a multiset $\mathcal{D}=\{\sigma_0,\ldots,\sigma_{w'-1}\}$.
In convention, the element $\sigma$ with multiplicity $m$ in $\mathcal{D}$ is denoted by $\sigma^m$, where the upper index is ommited when it is equal to $1$.
Fig.~\ref{Fig.1-different}  illustrates the 1-positions for the case of $L=15, T=3$, and  $\beta=15$. 
According to Definition~\ref{def:1-position},  the distribution of 1-positions is determined by the characteristic sets. We find that in some special cases, the distribution of 1-positions exhibits specific properties. We list the results for three special cases in Lemma~1 and relegate the proof in Appendix~\ref{appendix:lemma:1-position}.
Among them, the first result will play an important role in the derivation of low-complexity closed-form expression for the case of $\gcd(T,L)=1$. 
Although the properties for the other two cases will not be used in this paper's proofs, they may be useful in future discussions or analyses for the two special cases. 
By introducing all three cases together, we provide a more complete picture of the special patterns in 1-position distributions. 

\begin{lemma} \label{lemma:1-position}
The followings hold.
\begin{enumerate}[(i)]
    \item When $\gcd(T,L)=1$, $\mathcal{D}=\{0^w,1^w,\ldots,(T-1)^w\}$, for $\bm{s}_i \in \{\bm{v}_1, \bm{v}_2, \ldots, \bm{v}_{p+1}\}$.
    \item When $\gcd(T,L)=p$, 
$\mathcal{D}=\{0,1,\ldots,T-1\}$, for $\bm{s}_i \in \{\bm{v}_1, \bm{v}_2, \ldots, \bm{v}_{p+1}\}$.
    \item When  $T|q$, 
\begin{equation*} 
    \mathcal{D}=\begin{cases}
\{j \mod{T}:j\in \mathbb{Z}_{w}\}, \text{if } \bm{s}_i \in \{\bm{v}_1, \bm{v}_2, \ldots, \bm{v}_{p}\}, \\
\{\underbrace{0,0,\ldots,0}_{w}\}, \text{if } \bm{s}_i =  \bm{v}_{p+1}.
\end{cases}
\end{equation*}
    
\end{enumerate}
\end{lemma} 

\section{Analysis on AoI under MHUI Sequence Scheme} \label{sec:general results}


According to the CRT construction introduced in Section~\ref{sec:CRT sequence}, the length of MHUI sequences,  $L$, follows a quadratic order of complexity, i.e., $O(N^2)$. Consequently, calculating $\overline{A}$ by enumerating all cases of $\bm{\tau}$ based on \eqref{eq:AOI} leads to a significantly high complexity of $O(N^{2N}\rho)$.
In this section, we aim to derive a closed-form expression for $\overline{A}$ with lower computational complexity.
We  will develop a mathematical approach based on integer partitions to facilitate calculation.

\vspace{-0.2cm}
\subsection{Low-complexity Closed-form Expression for $\overline{A}$ under MHUI Sequence Scheme}


The key idea to obtain a low-complexity closed-form solution is to consider all possible transmission outcomes rather than enumerating all possible offset vectors. This strategy significantly narrows down the enumeration space, thereby greatly lowering the computational complexity. In the following, we introduce the derivation of low-complexity closed-form on $\overline{A}$ in detail.

\begin{figure}[t] 
	\centering 
\includegraphics[width=0.48\textwidth]{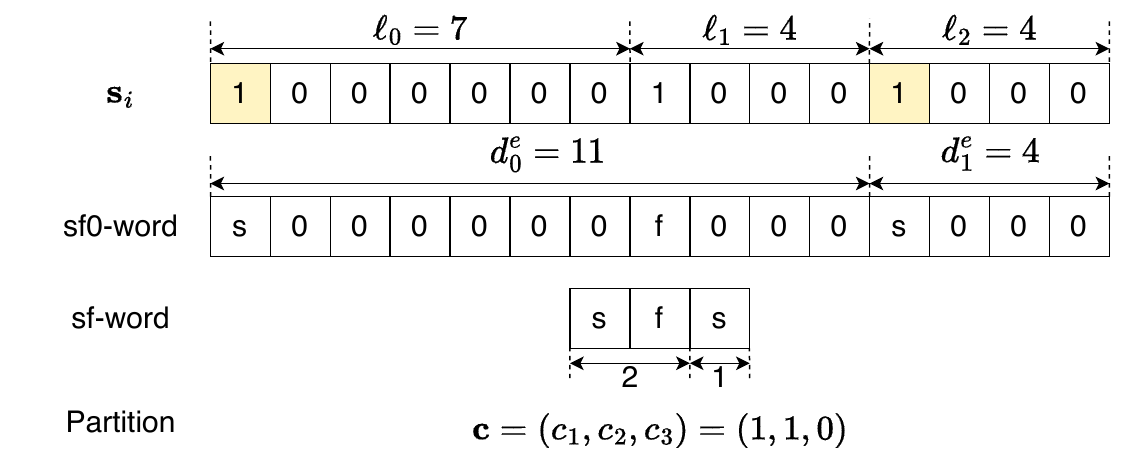} 
\vspace{-0.2cm}
\caption{Illustration of sequence $\bm{s}_i$, $\mathsf{sf0}$-word, $\mathsf{sf}$-word and an $r$-partition of $w$, and two classes of distances between two ``1''s,
$d^e_0,d^e_1,\ldots,d^e_{r-1}$ and $\ell_0, \ell_1, \ldots, \ell_{w-1}$. The sequence $\bm{s}_i$ used is $\bm{v}_2$ in \eqref{example_s2}, which is of Hamming weight $w=3$ and length $L=15$. Under the given event $e$, the number of successful ``1''s within $\bm{s}_i$ is $r=2$, and their positions  are marked in yellow. This event can be represented by the $\mathsf{sf}$-word $\mathsf{sfs}$, which can be mapped to the partition  $\bm{c}=(1,1,0)$ under the mapping $\theta$.
 }

	\label{Fig.illustration of ell and d}    \vspace{-0.2cm} 
\end{figure}

For any user~$i$, to derive $\overline{A}$ under the MHUI sequence scheme, we should consider the  AoI of user~$i$  across all possible  offset vectors, $\bm{\tau}$'s. Different offset vectors may result in different transmission outcomes, called \emph{events},  for the $w$ ``1''s in sequence $\bm{s}_i$. 
Consider an event where the number of successful ``1''s in $\bm{s}_i$ is equal to $r$, $r\in \{1,2,\ldots, w \}$.
We represent a successful ``1'' with a symbol $\mathsf{s}$ and a failed ``1'' with a symbol $\mathsf{f}$. As such, sequence $\bm{s}_i$ can be converted into an $\mathsf{sf0}$-word.
By removing all the ``0''s in the $\mathsf{sf0}$-word, we can further obtain an $\mathsf{sf}$-word of length $w$, which contains exactly $r\,\mathsf{s}$'s and $(w-r)\,\mathsf{f}\,$'s, as illustrated in Fig.~\ref{Fig.illustration of ell and d}. 
Denote by $\mathcal{E}_r$ the event set where each element represents an event with $r$ successful transmissions  within $\bm{s}_i$. Each element of $\mathcal{E}_r$ can be indexed by an $\mathsf{sf}$-word.
Obviously, $|\mathcal{E}_r|=\binom{w}{r}$.


For an event $e\in\mathcal{E}_r$, let $t^e_0 < t^e_1 < \cdots < t^e_{r-1}$ denote the positions of the $r$ successful transmissions within $\bm{s}_i$, and $d^e_k$ denote the \textit{cyclic distance} between the $k$-th and $(k\oplus_r 1)$-th successful transmissions within $\bm{s}_i$, for $k\in \mathbb{Z}_r $. That is, $d^e_k=t^e_{k+1}-t^e_k$ for $k=0,1,\ldots,r-2$, and $d^e_k=t^e_0+L-t^e_{r-1}$ for $k=r-1$.
We have $\sum_{k=0}^{r-1}d^e_k=L$.
It is worth noting that both the two classes of symbols $d^e_0,d^e_1,\ldots,d^e_{r-1}$ and $\ell_0,\ell_1,\ldots,\ell_{w-1}$ are used to describe cyclic distances between two ``1''s in $\bm{s}_i$. However, the former represents the distances between two adjacent successful ``1''s, which vary under different offset vectors and are unknown beforehand, while the latter is determined by the sequence construction method and thus is known beforehand. 
We illustrate the differences between them in Fig.~\ref{Fig.illustration of ell and d}.

Denote by $P_r$ the probability that an arbitrary event in $\mathcal{E}_r$ occurs over all offset vectors. 
Theorem~\ref{lemma:bound ES-P}  provides a closed-form expression for $P_r$ and its proof is given in Appendix~\ref{appendix:lemma:bound ES-P}. 

\begin{theorem}\label{lemma:bound ES-P}
For $r\in \{1,2,\ldots, w \}$, the probability for each event in {$\mathcal{E}_r$} to occur is given by \eqref{eq:p_t}. 
\end{theorem}  

\begin{figure*}
    \begin{equation} \label{eq:p_t}
P_r=\frac{1}{\binom{w}{r}L^{N-1}}\sum_{n=w-r}^{N-1}\sum_{m=0}^{w-r}(-1)^{m} \binom{N-1}{n}\binom{w-r}{m}(w-r-m)^{n}w^n (L-w^2)^{N-1-n}.
\end{equation}   
\end{figure*}

The computation complexity of $P_r$ in \eqref{eq:p_t} is $O(N^3)$. This is because each component inside the double summations is of complexity $O(N)$, and each summation contributes to operating times multiplying $O(N)$.
For an event $e\in\mathcal{E}_r$, we denote by $S^e_j$ the service time of the packet delivered at the $j$-th AoI drop, and $Y^e_j$ the slot-level inter-departure time interval between the $j$-th and $(j+1)$-th AoI drops.
Then, $\overline{A}$ in~\eqref{eq:AOI} can be calculated~as
\begin{align} \label{eq:A in S and Y}
   \overline{A} = \sum_{r=1}^w P_r\sum_{e\in\mathcal{E}_r}\left( \frac{\mathbb{E}_j[S^e_jY^e_j]}{\mathbb{E}_j[Y^e_j]}+\frac{\mathbb{E}_j[(Y^e_j)^2]}{2\mathbb{E}_j[Y^e_j]}-\frac{1}{2} \right). 
\end{align}
Furthermore, denote by $\varepsilon^e_j$ the end of the frame within which the $j$-th AoI drop of user~$i$ occurs. 
Now, we define $X^e_j$ as the time interval
between the ends of two frames where the  $j$-th and $(j+1)$-th AoI drops of user~$i$ occur,  i.e., $X^e_j=\varepsilon^e_{j+1} -\varepsilon^e_{j}$.
Compared to $Y^e_j$,  which represents the slot-level inter-departure time between two consecutive AoI drops, $X^e_j$ represents the \emph{frame-level inter-departure time} between two consecutive AoI drops, whose value is taken a multiple of $T$, as illustrated in Fig.~\ref{Fig.illustration of X and Y}.  
We have the following result on $Y^e_j$ and $X^e_j$. 
The proof is given in Appendix~\ref{appendix:lemma:EX=EY}. 

\begin{lemma} \label{lemma:EX=EY}
For $e\in\mathcal{E}_r$, we have $Y^e_j=X^e_j+S^e_{j+1}-S^e_j$.
Moreover, we have $\mathbb{E}_j[Y^e_j]=\mathbb{E}_j[X^e_j]$. 
\end{lemma}

Then  \eqref{eq:A in S and Y} can be rewritten, by Lemma~\ref{lemma:EX=EY}, as
\begin{gather} \label{eq:A in S and X}
  \overline{A} =\sum_{r=1}^w P_r\sum_{e\in\mathcal{E}_r} \left( \frac{\mathbb{E}_j[S^e_{j+1}X^e_j]}{\mathbb{E}_j[X^e_j]}+\frac{\mathbb{E}_j[(X^e_j)^2]}{2\mathbb{E}_j[X^e_j]}-\frac{1}{2} \right). 
\end{gather}

\begin{figure}[t] 
	\centering 
\includegraphics[width=0.48\textwidth]{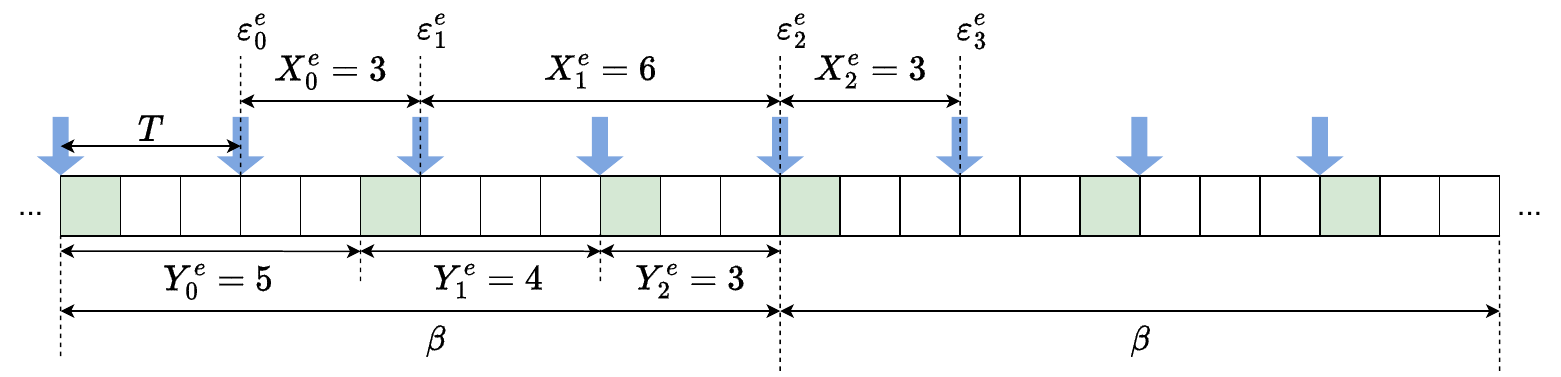} 

	\caption{Illustration of $Y_j^e$ and $X_j^e$.
 In the given scenario, $T=3$, $L=4$,  $\beta=12$.  Within a superframe, there are three AoI drops in total. The time slots that experience AoI drops are marked in green. The slot-level inter-departure times   between two consecutive drops within a superframe are 5,4,3,  respectively. 
 The frame-level inter-departure times   between two consecutive drops within a superframe are 3,6,3, respectively.
 }
	\label{Fig.illustration of X and Y}   

\end{figure}


Recall that for each $\bm{\tau}$, we use $O(\rho)$ to denote the computational complexity of $\overline{A_{\bm{\tau}}}$. The component inside the bracket in either \eqref{eq:A in S and Y} or \eqref{eq:A in S and X} calculates the average AoI for a given event $e$, which has the same computational complexity as $\overline{A_{\bm{\tau}}}$, that is, $O(\rho)$. This is because, no matter given $\bm{\tau}$ or given event $e$, calculating the average AoI requires considering the distributions of successful ``1''s and computing the statistics of the service times and inter-departure times accordingly. The double summations outside the bracket in either \eqref{eq:A in S and Y} or \eqref{eq:A in S and X} represent the calculation of average AoI under all possible  events. The number of all possible events is in the order of $2^N$. Consequently, both \eqref{eq:A in S and Y} and \eqref{eq:A in S and X} have a computational complexity of $O(2^{N} \rho)$, which is considerably lower than $O(N^{2N} \rho)$ by~\eqref{eq:AOI}. 

The two formulas involve different variables: \eqref{eq:A in S and Y} involves variable $Y_j^e$, which represents the slot-level inter-departure time between two consecutive AoI drops; \eqref{eq:A in S and X} involves variable $X_j^e$, representing the frame-level inter-departure time. 
 This allows for analysis from different perspectives.
By providing both forms, this paper offers flexibility for readers  to choose the most suitable approach for their specific cases. An upper bound on $\overline{A}$ can be obtained from either \eqref{eq:A in S and Y} or \eqref{eq:A in S and X}. As the sequence scheme can guarantee at least one AoI drop at a consistent location within each superframe, we have  $Y_j^e \leq \beta $ and $X_j^e \leq \beta $. By definition, we have  $S_j^e \leq T-1$. Therefore,  the upper bound $\overline{A} \leq T+ \frac{ \beta-3}{2}$ is derived.

\begin{remark}
We derived a closed-form formula for $ \overline{A}$ in our previous work \cite{liu2023Age} for the case of $T=L$. Note that that formula can be viewed as a special case of both  \eqref{eq:A in S and Y} and \eqref{eq:A in S and X}, in which  both  \eqref{eq:A in S and Y} and \eqref{eq:A in S and X} can be simplified as $ \overline{A} =\sum_{r=1}^w P_r\sum_{e\in\mathcal{E}_r} \left( \mathbb{E}_j [S^e_{j+1}]+\frac{L-1}{2} \right) $ with $Y_j^e = X_j^e =L$. This is because when $T=L$,  for any user, the drop of instantaneous AoI must happen in every sequence period, and the positions of its AoI drops within every sequence period  are the same. 
Consequently, for any user, the inter-departure time between any two consecutive drops under any event is given by $Y_j^e = X_j^e =L$.
\end{remark}

Both \eqref{eq:A in S and Y} and \eqref{eq:A in S and X} require tallying over all the possible choices of $d^e_0,d^e_1,\ldots,d^e_{r-1}$, 
and can be expressed as
$
  \overline{A} =\sum_{r=1}^w P_r \sum_{e\in\mathcal{E}_r} G(d^e_0, d^e_1,\ldots, d^e_{r-1}) ,
$
where $G(d^e_0, d^e_1,\ldots, d^e_{r-1})$ is a function of $d^e_0, d^e_1,\ldots, d^e_{r-1}$, representing the calculation for average AoI under event~$e$.
 Usually, $G(d^e_0, d^e_1,\ldots, d^e_{r-1})$ can be simplified to $\sum_{k=0}^{r-1} F(d^e_k)$, where $F(d^e_k)$ is a function of $d^e_k$. That is, 
\begin{gather} \label{eq:A in F}
  \overline{A} =\sum_{r=1}^w P_r \sum_{e\in\mathcal{E}_r} \sum_{k=0}^{r-1} F(d^e_k) .
\end{gather}
Based on this simpliﬁcation, we develop a mathematical tool built on the idea of integer partitions in the following subsection, to further reduce the complexity of \eqref{eq:A in F}.  With this, we can derive a formula for $\overline{A}$ that is directly based on the parameters $\ell_0,\ell_1,\ldots,\ell_{w-1}$. 
We will use two cases for illustration in Section~\ref{sec:specail cases}.

\vspace{-0.4cm}
\subsection{Time Statistics Calculation Based on Integer Partitions } \label{subsec:integer partition}

An efficient evaluation algorithm for the term $ \sum_{e\in\mathcal{E}_r} \sum_{k=0}^{r-1} F(d^e_k)$ in \eqref{eq:A in F} is to map each event $e\in\mathcal{E}_r$  to an integer partition of $w$. 
For two positive integers 
$(w,r)$,
an \emph{$r$-partition of $w$}
is a decomposition of $w$ so that
$w=\lambda_1+\lambda_2+\cdots+\lambda_r$, where each $\lambda_j\geq 1$ for all $j\in \{1,2,\ldots,r \}$.
Note that the ordering of $\lambda_j$'s does not matter.
The summands $\lambda_j$'s are called the \emph{parts} of $w$.

Denote by $\mathcal{P}^w_r$ the collection of all $r$-partitions of $w$. 
Now, define a mapping $\theta:\mathcal{E}_r\to\mathcal{P}^w_r$ by letting $\lambda_j$ be the cyclic distance of the positions of the $j$-th and $(j\oplus_w 1)$-th $\mathsf{s}$'s in an $\mathsf{sf}$-word.
Notice that the mapping $\theta$ is surjective, because for a given $r$-partition,
there is always a corresponding 
$\mathsf{sf}$-word: $ W_{SF}=
\mathsf{s}\underbrace{\mathsf{f}\cdots\mathsf{f}}_{\lambda_1-1}\mathsf{s}\underbrace{\mathsf{f}\cdots\mathsf{f}}_{\lambda_2-1}\cdots\mathsf{s}\underbrace{\mathsf{f}\cdots\mathsf{f}}_{\lambda_r-1}$.
Note that an $r$-partition of $w$ can be represented by a vector $\bm{c}=(c_1,c_2,\ldots,c_w)$, where $c_j$ denotes the number of part $j$.
That is, $\sum_{j=1}^wc_j=r$ and $\sum_{j=1}^wjc_j=w$. 
We illustrate the mapping between an $\mathsf{sf}$-word and an $r$-partition of $w$ in Fig.~\ref{Fig.illustration of ell and d}.

\begin{lemma} \label{lemma:partition-preimage}
For a given vector $\bm{c}=(c_1,c_2,\ldots,c_w) \in \mathcal{P}^w_r$, its pre-image set size is
\begin{equation}\label{eq:partition-preimage}
    |\theta^{-1}(\bm{c})|=\frac{(r-1)!w}{c_1!c_2!\cdots c_w!}.
\end{equation} 
\end{lemma}

We relegate the proof of Lemma~\ref{lemma:partition-preimage} in Appendix~\ref{appendix:parition-preimage}.

 Note that if we cyclically left shift an $\mathsf{sf}$-word, the resulting word also corresponds
to a valid event.  For example, cyclically left shifting 
$W_{SF}$ to 
$W^{'}_{SF}=\underbrace{\mathsf{f}\cdots\mathsf{f}}_{\lambda_1-1}\mathsf{s}\underbrace{\mathsf{f}\cdots\mathsf{f}}_{\lambda_2-1}\cdots\mathsf{s}\underbrace{\mathsf{f}\cdots\mathsf{f}}_{\lambda_r-1}\mathsf{s}$
corresponds to an event starting with 
${\lambda_1-1}$ failures before the first success.
Moreover, the sum $ \sum_{e\in\mathcal{E}_r} \sum_{k=0}^{r-1} F(d^e_k)$ has a special structure
related to these cyclic shifts.  Continuing with the
$W_{SF}$ example, if the corresponding event is
labelled as $e_1$, then 
$F(d^{e_1}_0)=F(\ell_0 \oplus_{w} \ldots \oplus_{w} \ell_{\lambda_1-1})$.
If $e_2$ is the event that corresponds to $W^{'}_{SF}$,
then
$F(d^{e_2}_{r-1})=F(\ell_{w-1} \oplus_{w} \ell_0 \oplus_{w} \ldots \oplus_{w} \ell_{\lambda_1-2} )$, and by means of
shifts with $i$ steps, where $0 \leq i <w$,
one can recover all terms of the form
$F(\ell_{i} \oplus_{w} \ell_{i+1} \oplus_{w} \ldots \oplus_{w} \ell_{\lambda_1+i-1} )$ in the sum.

Define
\begin{equation} \label{eq:b_k-e1}
\zeta_w(k,j)  \triangleq \sum_{i=0}^{j-1} \ell_{k\oplus_{w} i},~~~
b_j \triangleq \sum_{k=0}^{w-1} F(\zeta_w(k,j)) .
\end{equation}
(Note that $\zeta_w(k,w)=L$ for all $k$.)

It follows that
$ \sum_{e\in\mathcal{E}_r} \sum_{k=0}^{r-1} F(d^e_k)$ can be converted to an expression involving
$\{ b_1, \ldots, b_w \}$.
The exact relation is given by Lemma~\ref{lemma:bound ES-e3}, the proof of which  is given in Appendix~\ref{appendix:lemma:bound ES-e3}. 

\begin{lemma} \label{lemma:bound ES-e3} 
For sequence $\bm{s}_i$ with Hamming weight $w$, given the total number of successful ``1''s within $\bm{s}_i$, $r$, $r\in \{1,2,\ldots,w\}$, we have
\begin{equation} \label{eq:bound ES-e5}
\sum_{e\in\mathcal{E}_r} \sum_{k=0}^{r-1} F(d^e_k)  = \sum_{\bm{c} \in \mathcal{P}^w_r} \frac{|\theta^{-1}(\bm{c})|}{w} \sum_{j=1}^w c_j b_j .
\end{equation}
\end{lemma}

To evaluate the RHS of \eqref{eq:bound ES-e5}, we will construct a generating function $Q(x,y)$ as follows. We will demonstrate that the RHS of \eqref{eq:bound ES-e5} can be obtained by multiplying  the coefficient of $x^w y^r $ in $Q(x,y)$ by $(r-1)!$, as indicated in Lemma~\ref{lemma:c_q}. The closed-form expression of this coefficient is provided by Lemma~\ref{lemma:c_q2}.

\begin{lemma}\label{lemma:c_q}
  Let 
$g_j (x,y)=1+\frac{x^j y}{1!}+\frac{x^{2j} y^2}{2!}+ \frac{x^{3j} y^3}{3!}+ \cdots , 
 \Psi(x,y)=\prod_{j=1}^{\infty}g_j(x,y) , 
 Q(x,y)=y\Psi(x,y) \sum_{j=1}^{\infty} b_j x^j .$    
We denote  the coefficient of $x^w y^r$ in $Q(x,y)$ by $C_Q(w,r)$. Then we have 
\begin{equation}
\sum_{\bm{c} \in \mathcal{P}^w_r} \frac{|\theta^{-1}(\bm{c})|}{w} \sum_{j=1}^w c_j b_j= (r-1)! C_Q(w,r).   
\end{equation} 
\end{lemma}

\begin{lemma} \label{lemma:c_q2}
\begin{equation*}
 C_Q(w,r)=\begin{cases}   b_w,  \text{if } r=1, \\ \sum_{j=1}^w\frac{ (w-j-1)!b_j}{(r-1)!(r-2)!(w-j-(r-1))!},  \text{if } r\in \{2,\ldots,w\}.
 \end{cases}
\end{equation*} 
\end{lemma}

The proofs of Lemma~\ref{lemma:c_q} and Lemma~\ref{lemma:c_q2} are given in Appendix~\ref{appendix:c_q} and Appendix~\ref{appendix:c_q2}, respectively. 
By combining Lemmas~\ref{lemma:bound ES-e3}, \ref{lemma:c_q} and \ref{lemma:c_q2}, we have the following result in Theorem~\ref{lemma:F}. 
\begin{theorem} \label{lemma:F}
For sequence $\bm{s}_i$ with Hamming weight $w$, given the total number of successful ``1''s within $\bm{s}_i$, $r$, we have

\begin{equation}\label{eq:ES-bound-F-e1}
\sum_{e\in\mathcal{E}_r} \sum_{k=0}^{r-1} F(d^e_k) =\begin{cases}   b_w, \text{if } r=1, \\
 \sum_{j=1}^{w} \frac{ (w-j-1)!b_j}{(r-2)!(w-j-(r-1))!}, \text{if } r\in \{2,\ldots,w\},
 
\end{cases}
\end{equation}
 where $b_j$ is given by \eqref{eq:b_k-e1}, for $j\in \{1,2,\ldots,w \}$.
\end{theorem}

By plugging \eqref{eq:ES-bound-F-e1} into \eqref{eq:A in F}, we have
\begin{equation}
     \overline{A} =P_1 w F(L) + \sum_{r=2}^w P_r  \sum_{j=1}^{w} \frac{ (w-j-1)!\sum_{k=0}^{w-1} F(\zeta_w(k,j))}{(r-2)!(w-j-(r-1))!}.
\end{equation}

As observed, the idea of integer partitions  brings the complexity of $\overline{A}$ down to  $O(N \max(N\rho,N^3))$. 
The term $\sum_{k=0}^{w-1} F(\zeta_w(k,j))$ has a complexity of $O(\rho)$. Consequently, the inner summation over $j$ incurs a complexity of $O(N\rho)$.  To determine the overall complexity, we compare this with the complexity of $P_r$, $O(N^3)$, and take the maximum of these two values. The outer summation then contributes an additional complexity factor of $O(N)$. Thus, the overall complexity is the product of these factors, leading to $O(N \max(N\rho,N^3))$.


\section{Analysis for AoI under Sequence Scheme in Two Cases}
\label{sec:specail cases}
In this section, we will provide a more in-depth analysis for two cases under
the assumption $T<L$.
The first is $\gcd(T,L)=1$, and the second is when each frame contains at most one ``1''. We will derive low-complexity closed-form expressions for $\overline{A}$ in these two cases. 

\vspace{-0.2cm}
\subsection{Case 1: $\gcd(T,L)=1$}
\begin{lemma} \label{lemma:bound ES-e1}
When  $\gcd(T,L)=1$,  for an event $e\in\mathcal{E}_r$,  $r\in \{1,2,\ldots, w \}$,  let
  $f^e_k=d^e_k -\left\lfloor \frac{d^e_k}{T} \right\rfloor T -1$, $k\in \mathbb{Z}_r $, and define $F_1(d^e_k)$ as in \eqref{eq:F_1_d^e_k},
then the following holds:   
\begin{gather} \label{eq:F1-case1}
    \frac{\mathbb{E}_j[S^e_{j+1}X^e_j]}{\mathbb{E}_j[X^e_j]}+\frac{\mathbb{E}_j[(X^e_j)^2]}{2\mathbb{E}_j[X^e_j]}-\frac{1}{2}=\sum_{k=0}^{r-1} F_1(d^e_k) .
\end{gather}
\end{lemma} 

\begin{figure*}
    \begin{gather} \label{eq:F_1_d^e_k}
    F_1(d^e_k)=\frac{\left \lceil \frac{d^e_k}{T} \right \rceil   f^e_k(f^e_k+1) + \left \lfloor \frac{d^e_k}{T} \right \rfloor  (f^e_k+T)(T-f^e_k-1)+\left( \left \lfloor \frac{d^e_k}{ T} \right \rfloor (2d^e_k -T) - \left \lfloor \frac{d^e_k}{ T} \right \rfloor^2 T +d^e_k \right) T}{2L} - \frac{1}{2}. 
\end{gather}
\end{figure*}

The proof of Lemma~\ref{lemma:bound ES-e1}  is given in Appendix~\ref{appendix:lemma:bound ES-e1}. Based on Lemma~\ref{lemma:bound ES-e1} and the tool of integer partitions, the low-complexity closed-form expression of $\overline{A}$ for the case where $\gcd(T,L)=1$ is obtained in Theorem~\ref{theorem:closed-form-coprime}.

\begin{theorem} \label{theorem:closed-form-coprime} 
When  $\gcd(T,L)=1$, given $\{\ell_0,\ell_1,\ldots,\ell_{w-1}\}$, 
 the following holds:
  \begin{gather}
  \overline{A} =P_1 w F_1(L)+ \sum_{r=2}^w P_r  \sum_{j=1}^{w} \frac{ (w-j-1)!\sum_{k=0}^{w-1} F_1(\zeta_w(k,j))}{(r-2)!(w-j-(r-1))!},
  \label{eq:closed-form for A case 1}
 \end{gather}
 where $P_r$ is given by \eqref{eq:p_t}.
\end{theorem} 

\vspace{-0.2cm}
\subsection{Case 2:  Each Frame Contains at Most One ``1''}
 \label{subsection:S for special case 2}

Here we consider the case where there is at most one ``1'' within each frame. Two examples for this case are provided in Lemma~\ref{lemma:case2-1} and Lemma~\ref{lemma:case2-2} below. Their proofs are given in Appendix~\ref{appendix:lemma:case2-1} and Appendix~\ref{appendix:lemma:case2-2}, respectively. 

\begin{lemma} \label{lemma:case2-1}
    Given $\bm{s}_i \in \{\bm{v}_2,\bm{v}_3,\ldots,\bm{v}_{p+1} \}$, if $T \leq p$, then each frame of user~$i$ contains at most one ``1''. 
\end{lemma}

\begin{lemma} \label{lemma:case2-2}
  Given $\bm{s}_i \in \{\bm{v}_2,\bm{v}_3,\ldots,\bm{v}_{p+1} \}$, if    $T=q$,  then each frame of user~$i$ contains at most one ``1''. 
\end{lemma}

In the case where each frame contains at most one ``1'', each status packet will not be transmitted more than once. This indicates that each successful ``1'' is corresponding to an AoI drop. Based on this observation, we obtain the following result. The proof  is given in Appendix~\ref{appendix:lemma:special case 2}.

\begin{lemma} \label{lemma:A-case2-1}
When each frame contains at most one ``1'', given $\{\ell_0,\ell_1,\ldots,\ell_{w'-1}\}$ and $\mathcal{D}=\{\sigma_0,\sigma_1,\ldots,\sigma_{w'-1}\}$,
then the following holds:
\begin{gather*} 
\sum_{e\in\mathcal{E}_r} \frac{\mathbb{E}_j[(Y^e_j)^2]}{2\mathbb{E}_j[Y^e_j]} =\sum_{e\in\mathcal{E}_r} \frac{\sum_{k=0}^{r-1}(d^e_k)^2}{2L} , \\
\sum_{e\in\mathcal{E}_r}\frac{\mathbb{E}_j[S^e_jY^e_j]}{\mathbb{E}_j[Y^e_j]} 
 =\sum_{\bm{c} \in \mathcal{P}^w_r}   \frac{|\theta^{-1}(\bm{c})|}{\beta w}  \sum_{j=1}^w c_j   \sum_{k=0}^{w'-1}\zeta_{w'}(k,j) \sigma_k .
\end{gather*}
\end{lemma} 

Based on Lemma~\ref{lemma:A-case2-1} and the tool of integer partitions, a low-complexity closed-form expression of $\overline{A}$ for this case is obtained in Theorem~\ref{theorem:case2-closed form}.

\begin{theorem} \label{theorem:case2-closed form}
When each frame contains at most one ``1'', given $\{\ell_0,\ell_1,\ldots,\ell_{w'-1}\}$ and $\mathcal{D}=\{\sigma_0,\sigma_1,\ldots,\sigma_{w'-1}\}$,
let  
$
F_2(\zeta_{w'}(k,j))= \frac{\zeta_{w'}(k,j) \sigma_k}{\beta}, 
F_3(\zeta_{w}(k,j))=\frac{(\zeta_w(k,j))^2}{2L}.
$
Then \eqref{eq:closed-form for A case 2} holds, where $P_r$ is given by \eqref{eq:p_t}.
\end{theorem}

\begin{figure*}
\begin{equation} \label{eq:closed-form for A case 2}
    \overline{A}=P_1 (w'F_3(\beta) +w F_3(L))+\sum_{r=2}^w P_r \sum_{j=1}^{w}  \frac{(w-j-1)!\left(\sum_{k=0}^{w'-1} F_2(\zeta_{w'}(k,j))+ \sum_{k=0}^{w-1} F_3(\zeta_{w}(k,j))\right) }{(r-2)!(w-j-(r-1))!}.
\end{equation}
\end{figure*}

One can check that the closed-form expressions in \eqref{eq:closed-form for A case 1} and \eqref{eq:closed-form for A case 2} for $\overline{A}$ in the two cases enjoy a polynomial complexity,  $O(N^4)$.
This is because the calculation of $ F_1(\zeta_w(k,j))$, $F_2(\zeta_{w'}(k,j))$ or $F_3(\zeta_{w}(k,j))$ is of complexity $O(N)$, and each of the triple summations in either \eqref{eq:closed-form for A case 1} or \eqref{eq:closed-form for A case 2} contributes to operating times multiplying $O(N)$. 
The complexity $O(N^4)$ is significantly lower than the exponential complexity of both \eqref{eq:A in S and Y} and \eqref{eq:A in S and X}, $O(2^{N} \rho)$.  
This reduction in complexity makes the closed-form expressions more efficient and practical for calculations, particularly when dealing with large values of~$N$.

\section{ Parameter Selection for Sequence Scheme} \label{sec:AoI optimization}
In this section, we consider AoI optimization under the sequence  scheme based on the analytical results obtained in previous sections. 
These  results provide important guidance for parameter selection in CRT construction. 
To obtain the MHUI sequences by CRT construction, three parameters need to be set: $p$, $q$, and $w$. Given $N$, we set $p$ as the smallest prime number that is no less than $N$, and $w$ as equal to $p$. As for the parameter $q$, it should satisfy two conditions: it should be no less than $(2p-1)$, and it should be coprime with $p$. Usually, we set $q=2p-1$ to achieve the shortest sequence length $L$.

However, it is important to note that the shortest $L$ may not necessarily correspond to the smallest $\overline{A}$. We have checked the numerical values of the closed-form expressions obtained for the two cases where $\gcd(T,L)=1$ and where each frame contains at most one ``1'', and found that $\overline{A}$ under $q=T$ is usually lower than that under $q=2p-1$. Based on this observation, we propose a heuristic algorithm for the selection of $q$ as follows.
Given $N$ and $T$, if $T\geq 2p-1$ and $\gcd(T,p(2p-1))=1$, the $\overline{A}$  under $q=2p-1$ can be derived by \eqref{eq:closed-form for A case 1}. If the resulting value is higher than $\overline{A}$  under $q=T$, which can be derived by \eqref{eq:closed-form for A case 2}, we then let $q=T$ instead of $q=2p-1$, and choose $\{\bm{s}_1, \bm{s}_2, \ldots, \bm{s}_N\}$ from $\{\bm{v}_2,\bm{v}_3,\ldots,\bm{v}_{p+1} \}$. Otherwise, we let $q=2p-1$, and choose $\{\bm{s}_1, \bm{s}_2, \ldots, \bm{s}_N\}$ from $\{\bm{v}_1,\bm{v}_2,\ldots,\bm{v}_{p+1} \}$.  We present the above heuristic parameter selection method in Algorithm~\ref{alg:CRT parameter}.

\begin{algorithm}[htb]
\caption{Algorithm for parameter selection in CRT construction}
\label{alg:CRT parameter}
\begin{algorithmic}[1] 
\REQUIRE System Parameters: $N, T$
\STATE $p \gets \text{The smallest prime number that satisfies } p \geq N$
\STATE $w \gets p$
    \IF{$T\geq 2p-1$ and $ \gcd(T,p(2p-1))=1$ }
            \STATE $A_1  \gets \overline{A} \text{ based on } (19) \text{ with 
             } q=2p-1$
            \STATE $A_2  \gets  \overline{A} \text{ based on } (20) \text{ with 
             } q=T$
                \IF{$A_1 > A_2$}
                \STATE $q=T$, choose $\{\bm{s}_1, \bm{s}_2, \ldots, \bm{s}_N\}$ from $\{\bm{v}_2,\bm{v}_3,\ldots,\bm{v}_{p+1} \}$
                \ELSE
                \STATE $q=2p-1$, choose $\{\bm{s}_1, \bm{s}_2, \ldots, \bm{s}_N\}$ from $\{\bm{v}_1,\bm{v}_2,\ldots,\bm{v}_{p+1} \}$
                \ENDIF
    \ENDIF
    
\ENSURE Parameters for CRT construction: $p,q,w$
\end{algorithmic}
\end{algorithm} 

Choosing $q=T$ instead of $q=2p-1$ is also helpful for reducing power consumption of the sequence scheme. We measure  power consumption by  duty factor, which is denoted by $f_s$ under the sequence scheme and is defined as the ratio of transmission slots within a time duration, that is, $f_s= \frac{w}{L}=\frac{1}{q}$. A higher duty factor implies higher energy consumption. 
Choosing $q=T$ with $T \geq 2N-1$ results in lower duty factor compared with choosing $q=2N-1$.
This indicates that by choosing $q=T$, the power efficiency of the sequence scheme can be  enhanced.

\begin{table*}[htbp]
\centering 
\caption{$N=7, T \in \{ 20,30,40,50,60 \} $}
\label{Table:  case 1 and case 2-2}
    \vspace{-0.2cm}
    \scalebox{0.95}{
\begin{tabular}{|c|c|c|c|c|c|} 
\hline
    $T$  &  20 & 30 & 40  & 50  & 60 \\
    \hline
\multirow{2}*{$q=2p-1$}  & 
$\overline{A} =22.78$  & $\overline{A} = 27.78 $ & $\overline{A} = 32.78 $  & $\overline{A} = 37.78 $  & $\overline{A} = 42.78$ \\
\cline{2-6} ~ &
$f_s=\frac{1}{13} $ & $f_s=\frac{1}{13} $ & $f_s=\frac{1}{13} $ & $f_s=\frac{1}{13} $ & $f_s=\frac{1}{13} $  \\
    \hline
\multirow{2}*{$q=T$} &  
$\overline{A} = 19.3$  & $\overline{A} = 24.02 $ & $\overline{A} = 28.89 $ & $\overline{A} = 33.81$  & $\overline{A} = 38.76$ \\
\cline{2-6} ~ &
$f_s=\frac{1}{20} $ & $f_s=\frac{1}{30}$ & $f_s=\frac{1}{40} $ & $f_s=\frac{1}{50} $ & $f_s=\frac{1}{60} $  \\
     \hline
\end{tabular}}
 \vspace{-0.2cm}
\end{table*}
We use the example of $N=7, T\in \{ 20,30,40,50,60 \}$  for illustration in Table~\ref{Table:  case 1 and case 2-2}.  According to Algorithm~1, we set $p=7$ and $w=7$. 
For each  $T$, we have $T\geq 2p-1,  \text{and}  \gcd(T,p(2p-1))=1$. 
Two choices of $q$, $q=2p-1$ and $q=T$, are considered. No matter under $q=2p-1$ or $q=T$, we let $\{ \bm{s}_1, \ldots, \bm{s}_7 \} = \{ \bm{v}_2, \ldots, \bm{v}_8  \}$. The $\overline{A}$ of  user~1, which is associated with $\bm{s}_1$, is shown in Table~\ref{Table:  case 1 and case 2-2}.  For $q=2p-1$, we derive  $\overline{A}$  by \eqref{eq:closed-form for A case 1}. For $q=T$, we  derive   $\overline{A}$  by  \eqref{eq:closed-form for A case 2}.
As observed, choosing $q=T$ leads to smaller  $\overline{A}$ compared to $q=2p-1$.
We have also shown the duty factors of  user~1 under  $q=2p-1$ and $q=T$ in Table~\ref{Table: case 1 and case 2-2}.
For example, when $T=60$,  the duty factor under $q=T$, $\frac{1}{60}$, is nearly 4.6x lower than that under $q=2p-1$, $\frac{1}{13}$. That is, the energy efficiency under $q=T$ is 4.6x higher than that under $q=2p-1$.

\begin{remark}
Note that in our previous work \cite{liu2023Age}, we also proposed sequence optimization method to improve AoI performance.   The optimization method in \cite{liu2023Age} and  this work are complementary.   
In this  work, our focus is on optimizing parameter selection before sequence construction. By carefully selecting appropriate parameters, especially for $q$, we can lay a solid foundation for constructing effective sequences that minimize AoI.
While in \cite{liu2023Age}, we concentrated on selecting the optimal cyclically shifted versions after sequence construction. By analyzing the performance of different shifted versions and choosing the optimal one, we further improve the sequence scheme's performance.
Both layers of optimization, parameter selection, and shifted version selection, are crucial for achieving optimal AoI performance under the sequence scheme. 
\end{remark}

\begin{remark}
Note that given $N$ and $T$, the shortest sequence length $L$, obtained by selecting $q=2p-1$, is independent of $T$. Specifically, this shortest sequence length $L=p(2p-1)$ is quadratic in $N$ since $p$ falls within the range  $N\leq p \leq 2N-1$ according to Bertrand's Postulate. Nevertheless, this $L$ may not yield the optimal AoI performance.
To enhance AoI performance, Algorithm~\ref{alg:CRT parameter} is proposed, which takes both parameters $N$ and $T$ into consideration simultaneously.
 When $T$ meets specific conditions (as indicated in line~3 in Algorithm~\ref{alg:CRT parameter}), the $\overline{A}$ under $q=T$ can be superior to that under $q=2p-1$. By letting $q=T$, a better choice of $L$ is given by $L=pT$, falling within the range  $NT \leq L \leq (2N-1)T$. 
This strategic modification enables AoI optimization by accounting for both the parameters $N$ and~$T$.
\end{remark}

\section{Simulation Results} \label{sec:simulation}
In this section, we present simulation results on AoI performance. We at first validate our derived analytical results by simulations, and then  compare  our proposed MHUI sequence scheme with two baseline schemes, slotted ALOHA and framed ALOHA. The two baselines are widely-used and well-studied distributed schemes in wireless communication systems. They provide a basic level of collision avoidance and are simple for implementation. 
We also compare their energy consumption based on their duty factors.
The duty factor of a scheme is defined as the ratio of transmission slots within a frame and indicates energy consumption.
 It is desirable for a scheme to achieve superior AoI performance while consuming less energy.
For each simulation experiment, we simulate the evolution of instantaneous AoI in $10^6$ runs to obtain the average $\overline{A}$ for all users. The starting time offsets of the users are randomly generated in each run.

\subsection{Comparison of Analytical Results against Simulation Results for MHUI Sequences} \label{subsec:simu-analytic}

\begin{figure}[t] 
  \begin{minipage}[b]{0.45\textwidth} 
    \centering
    \includegraphics[width=1\textwidth]{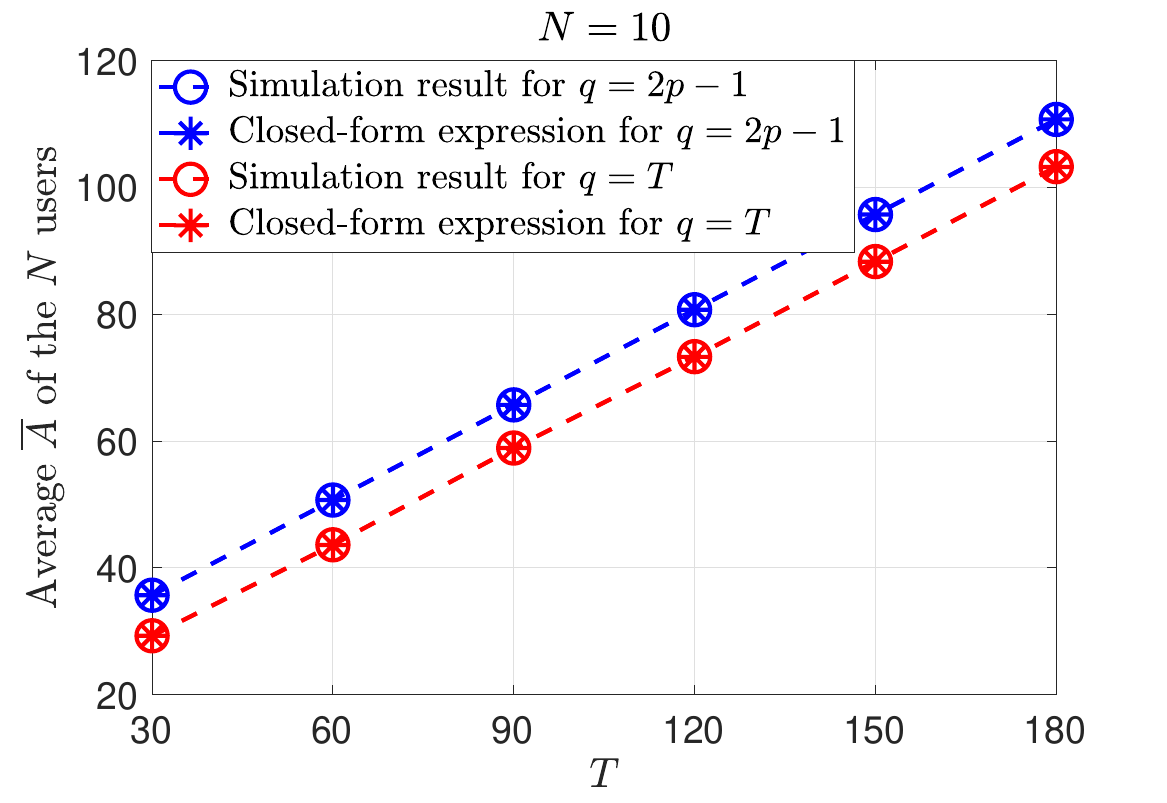}
	\caption{AoI performance in the case of $N=10$, $T\in \{30,60,90,120,150,180\}$ under the sequence scheme.}
   \label{Fig: N10_simu_theo_gap} 
  \end{minipage}%
  \hspace{0.15cm}
   \begin{minipage}[b]{0.5\textwidth} 
    \centering
    \scalebox{1}{
    \begin{tabular}{|c|c|c|c|c|c|c|} \hline 
         $T$&  30&  60&  90&  120&  150& 180\\ \hline 
         $q=2p-1$ &  $\frac{1}{21}$&  $\frac{1}{21}$&  $\frac{1}{21}$&  $\frac{1}{21}$&  $\frac{1}{21}$& $\frac{1}{21}$\\ \hline 
         $q=T$&  $\frac{1}{30}$&  $\frac{1}{60}$&  $\frac{1}{90}$&  $\frac{1}{120}$&  $\frac{1}{150}$& $\frac{1}{180}$\\ \hline 
    \end{tabular}}
    \tabcaption{The duty factor $f_s$ with different choices of $q$ under the sequence scheme} 
    \label{Table: duty factor N=10}
  \end{minipage}%
  \vspace{-1.0cm}
\end{figure}

We have derived analytical results  for AoI under  the MHUI sequence scheme. 
Here, we  compare them with simulation results.
We use the case of $N=10, T\in \{30,60,90,120,150,180 \}$ for illustration in Fig.~\ref{Fig: N10_simu_theo_gap}. 
For given $N$,  we have $p=11, w=11$ according to CRT construction. 
As each value of $T$ satisfies $T\geq 2p-1 $ and $\gcd(T,p(2p-1))=1$,  we consider two choices of $q$,  $q=2p-1$ and $q=T$. 
With $q=2p-1$, we have  $\gcd(T,L)=1$ for each $T$, and then the closed-form expression for $\overline{A}$ is given by Theorem~\ref{theorem:closed-form-coprime}. 
With $q=T$, each frame contains at most one ``1'', and then the closed-form expression for $\overline{A}$ is given by Theorem~\ref{theorem:case2-closed form}. 

We can observe from Fig.~\ref{Fig: N10_simu_theo_gap} that the closed-form expressions for $\overline{A}$ align well with the simulation results.  Moreover, as shown in Fig.~\ref{Fig: N10_simu_theo_gap},  the $\overline{A}$ under $q=T$ is lower than that under $q=2p-1$. For example, when $T=30$, the $\overline{A}$ under $q=T$ is equal to 29.3, which is $17.9\%$ lower than that under $q=2p-1$, 35.7. 
The duty factors under $q=2p-1$ and $q=T$ are listed in Table~\ref{Table: duty factor N=10}.
It is also noticeable that the duty factor under $q=T$ is lower than that under $q=2p-1$. In the example of $T=60$, the duty factor under $q=T$ is given by $\frac{1}{60}$, which is nearly $\frac{1}{3}$ of that under $q=2p-1$, $\frac{1}{21}$. 

\begin{figure}[h!]
\centering
\includegraphics[width=0.45\textwidth]{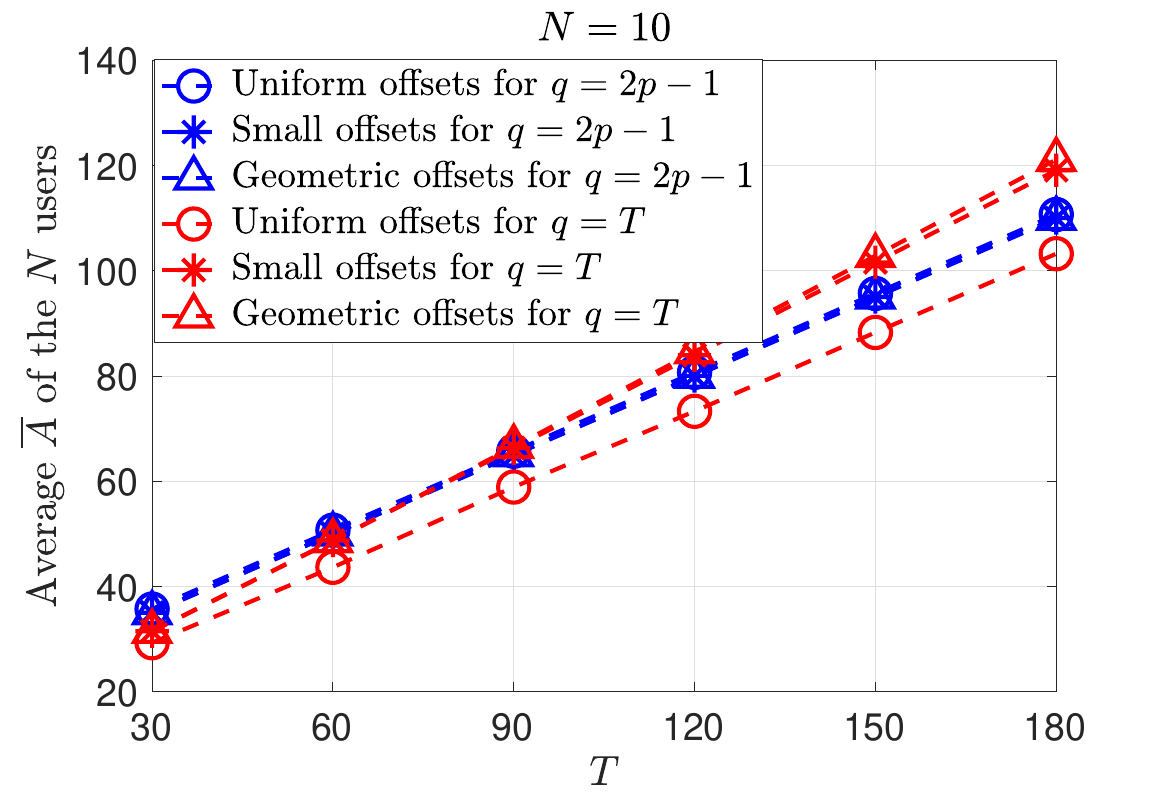}
\caption{AoI performance with various offset distributions in the case of $N=10$, $T\in \{30,60,90,120,150,180\}$ under the sequence scheme.}
 \label{Fig: non-uniform distribution}
\end{figure}

Note that the analytical results we derive in this paper are based on the assumption that the starting time offsets of the users are uniformly distributed. This assumption implies that each event in $\mathcal{E}_r$ occurs with  equal probability, which enables us to utilize the tool of integer partitions to achieve complexity reduction.  
If the offsets follow a different distribution such that each event in $\mathcal{E}_r$ has varying probabilities, then our current analytical results would not directly apply.  
We have conducted simulations to show the AoI performance of MHUI sequences under two scenarios of non-uniform distribution: 1)  We assume that the offsets are relatively small in comparison to the sequence period 
$L$, owing to a coarse level of time synchronization. Specifically, we assume that offsets are randomly selected from the interval $[0,L/4]$; 2) We consider the offsets to follow a geometric distribution, a commonly encountered distribution for discrete variables. For each user, the probability that a given slot is the offset termination slot is set at 0.01. We plot AoI performance under the two non-uniform distribution cases along with the uniform case in Fig.~\ref{Fig: non-uniform distribution} for MHUI sequences with two choices of $q$, $q=2p-1$ and $q=T$. As observed, for MHUI sequences with $q=2p-1$, the AoI performances under the two non-uniform distribution cases are close to that under uniform distribution case. While for MHUI sequences with $q=T$, the performance degradation is more obvious. 
It is also important to note that for systems characterized by a specific offset distribution instead of uniform distribution, there might be more suitable sequences than MHUI sequences. 


\subsection{Comparison between Sequence Scheme and the Baselines} 

Here we present simulation results comparing  AoI performance under the MHUI sequence scheme and the two baseline schemes, slotted ALOHA and framed ALOHA.  In the slotted ALOHA scheme, each user at each time slot transmits a packet with a fixed probability $p_t$ and remains idle with probability $(1-p_t)$. In the framed ALOHA scheme, each user randomly selects $w_{FA}$ time slots at the beginning of each frame to transmit packets, while keeping idle in the remaining slots, where $w_{FA}\in \{1,2,\ldots,T \}$. 

For the sequence scheme, the sequences employed are obtained by our proposed optimization algorithm. For the two baseline schemes, we optimize them separately. For the slotted ALOHA scheme, the optimal transmitting probability $p_t^*$ is given by \cite{yates2017status}: $p_t^*=1/N$. For the framed ALOHA scheme,  we find the optimal number of transmission attempts for each packet,  denoted by $w^*_{FA}$,  through an exhaustive search in the range $[1,T]$.
We also consider the duty factor of each scheme. The duty factor of each user in the  sequence scheme is given by $f_s= \frac{1}{q}$. Note that the duty factors of the optimal slotted ALOHA and  framed ALOHA schemes, denoted by $f_{SA}^*$ and $f_{FA}^*$, are usually higher than $f_s$. Specifically, $f_{SA}^*=1/N$, $f_{FA}^*=w^*_{FA}/T$. To make the comparison more complete, we also study the performance of the two baseline schemes with duty factor same as $f_s$ 
in parallel. 

\begin{figure}[t] 
  \begin{minipage}[b]{0.5\textwidth} 
    \centering
    \includegraphics[width=1\textwidth]{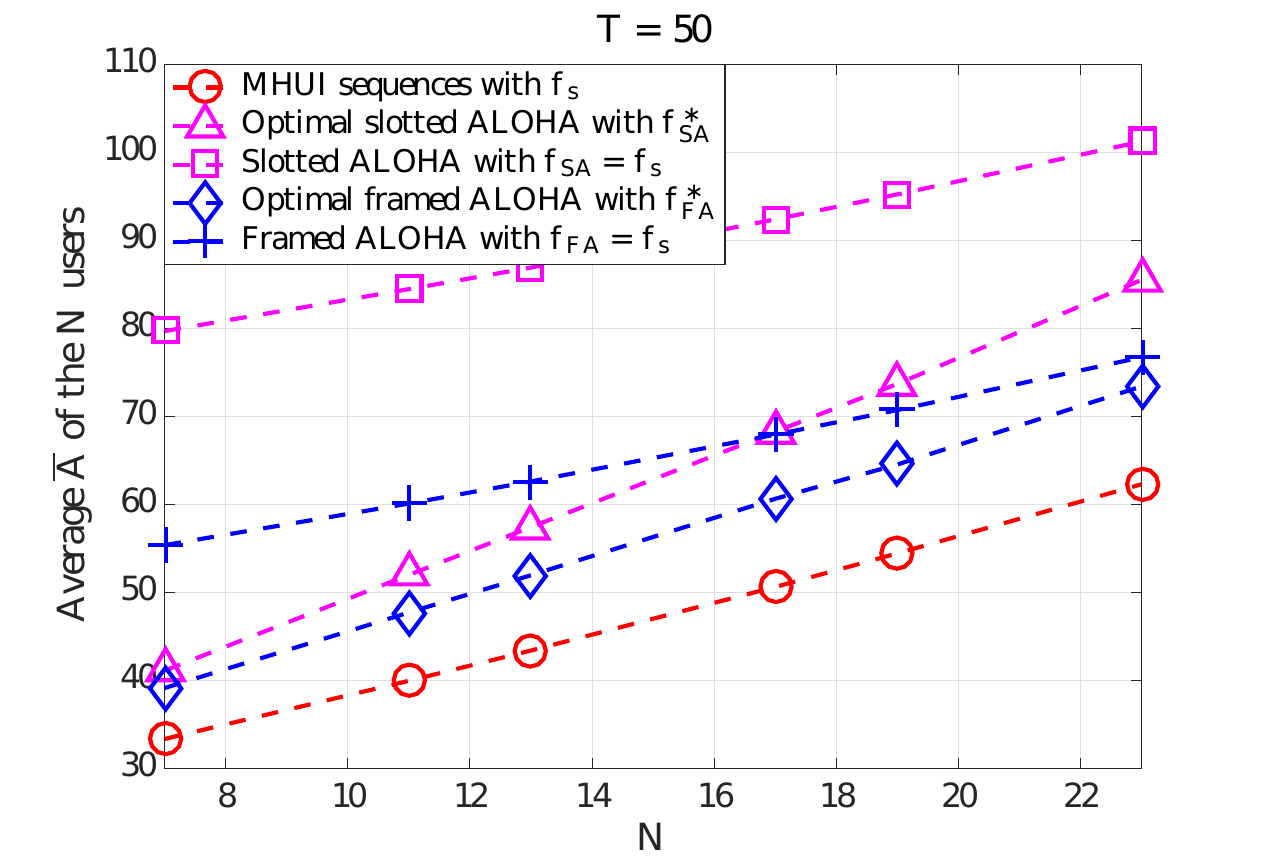}
	\caption{AoI performance under various schemes with $T=50$, $(N,L) \in \{(7,350),(11,550),(13,650),(17,850),(19,950),(23,1150)\}$.}
    \label{Fig: Simulation T=50} 
  \end{minipage}%
  \hspace{0.3cm}
   \begin{minipage}[b]{0.5\textwidth} 
    \centering
    \includegraphics[width=1\textwidth]{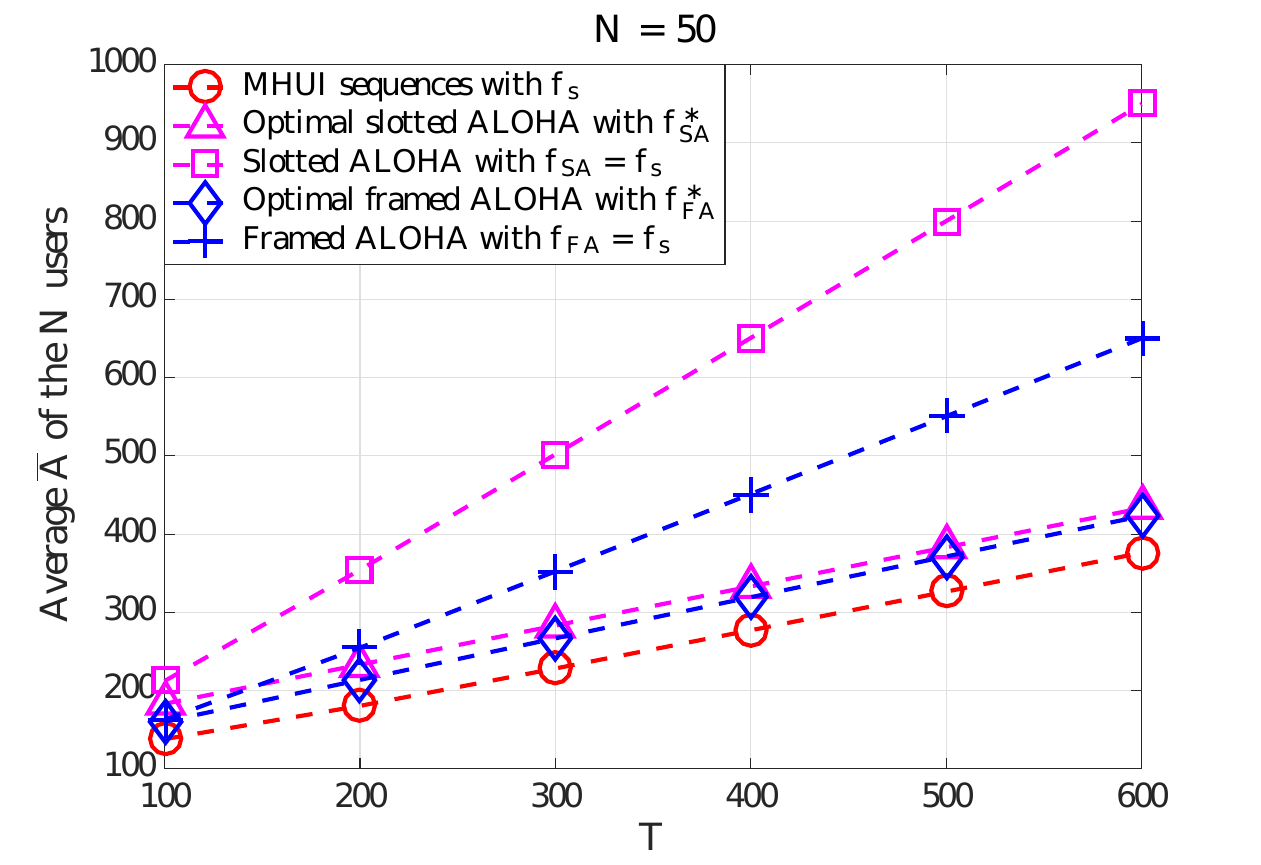}
	\caption{AoI performance under various schemes with $N=50$, $(T,L)\in\{(100,5300),(200,10600),(300,15900),(400,21200),(500,26500),\\ (600,31800)\}$.}
    \label{Fig: Simulation N=50} 
  \end{minipage}%
\end{figure}

We  compare 
 AoI performance of the  schemes when $T$ is fixed and $N$ is changing.
We show the case where $T=50$ and $N $  takes value from the prime number set $  \{7,11,13,17,19,23\}$ in Fig.~\ref{Fig: Simulation T=50}.    As shown, the sequence scheme achieves the shortest $\overline{A}$. For example, when $N=7$,  $\overline{A}$ under the sequence scheme is  33.38, which is 18.86\% lower than that under the optimal framed ALOHA scheme, 41.14. We show the duty factors of the schemes in Table~\ref{Table: duty factor T50}. 
Note that  $f_{FA}^*$ is close to $f_{SA}^*$, and both of 
them are higher than $f_s$. This implies that  employing either   the optimal slotted ALOHA scheme or the optimal framed ALOHA scheme costs more energy  than 
the  sequence  scheme.  For example,
in the case where $N=7$,  $f_s=\frac{1}{50}$, which is nearly $\frac{1}{7}$ of either $f_{FA}^*$ or $f_{SA}^*$. Overall, the sequence scheme enjoys enhanced AoI performance and achieves higher energy efficiency at the same~time. 

We also simulate the case where $N$ is fixed and $T$ is changing. Fig.~\ref{Fig: Simulation N=50} shows the comparison results on $\overline{A}$ when $N=50$ and $T\in \{100,200,300,400,500,600 \}$, and Table~\ref{Table: duty factor N50} shows the duty factors of the schemes. The results lead to the same conclusion that the MHUI sequence scheme attains superior AoI while improving energy efficiency, compared with the optimized baseline schemes. For example, when $T=300$, $\overline{A}$ under the MHUI sequence scheme is equal to 228.15, which is $14.4\%$ lower than that under the optimal framed ALOHA scheme, 266.68. At the same time, we have $f_s=\frac{1}{300}$, $f_{FA}^*=\frac{1}{50}$. That is, the MHUI sequence scheme enhances the energy efficiency by  6x compared with the optimal framed ALOHA scheme.

\begin{remark}
It is important to note that under the sequence scheme, the AoI performance of the users may vary, unlike the baselines. For instance, in the scenario where $N=50$ and $T=500$, the $\overline{A}$ of the user assigned with $\bm{v}_{p+1}$ is 300.8, while the $\overline{A}$ of the other $(N-1)$ users is 327. This discrepancy in AoI performance highlights the flexibility of the sequence scheme to accommodate users with varying demands for freshness of information.    
\end{remark}

\begin{figure}[t] 
  \begin{minipage}[b]{0.5\textwidth} 
    \centering
    \scalebox{0.85}{
    \begin{tabular}{|c|c|c|c|c|c|c|} \hline 
         $N$&  7&  11&  13&  17&  19& 23\\ \hline 
         $f_s$ &  $\frac{1}{50}$&  $\frac{1}{50}$&  $\frac{1}{50}$&  $\frac{1}{50}$&  $\frac{1}{50}$& $\frac{1}{50}$\\ \hline 
         $f_{SA}^*$&  $\frac{1}{7}$&  $\frac{1}{11}$&  $\frac{1}{13}$&  $\frac{1}{17}$&  $\frac{1}{19}$& $\frac{1}{23}$\\ \hline 
         $f_{FA}^*$&  $\frac{7}{50}$&  $\frac{2}{25}$&  $\frac{2}{25}$&  $\frac{3}{50}$&  $\frac{1}{25}$& $\frac{1}{25}$\\ \hline
    \end{tabular}}
    \tabcaption{The duty factors of various schemes under different $N$ with $T=50$} 
    \label{Table: duty factor T50}
  \end{minipage} 
    \begin{minipage}[b]{0.5\textwidth} 
    \centering
    \scalebox{0.85}{
    \begin{tabular}{|c|c|c|c|c|c|c|} \hline 
         $T$&  100&  200&  300&  400&  500& 600\\ \hline 
         $f_s$ &  $\frac{1}{100}$&  $\frac{1}{200}$&  $\frac{1}{300}$&  $\frac{1}{400}$&  $\frac{1}{500}$& $\frac{1}{600}$\\ \hline 
         $f_{SA}^*$&  $\frac{1}{50}$&  $\frac{1}{50}$&  $\frac{1}{50}$&  $\frac{1}{50}$&  $\frac{1}{50}$& $\frac{1}{50}$\\ \hline 
         $f_{FA}^*$&  $\frac{1}{50}$&  $\frac{1}{50}$&  $\frac{1}{50}$&  $\frac{1}{50}$&  $\frac{1}{50}$& $\frac{1}{25}$\\ \hline
    \end{tabular}}
    \tabcaption{The duty factors of various schemes under different $T$ with $N=50$} 
    \label{Table: duty factor N50}
  \end{minipage} 
 \vspace{-1cm}
\end{figure}

\begin{figure}[H]
\centering
\includegraphics[width=0.5\textwidth]{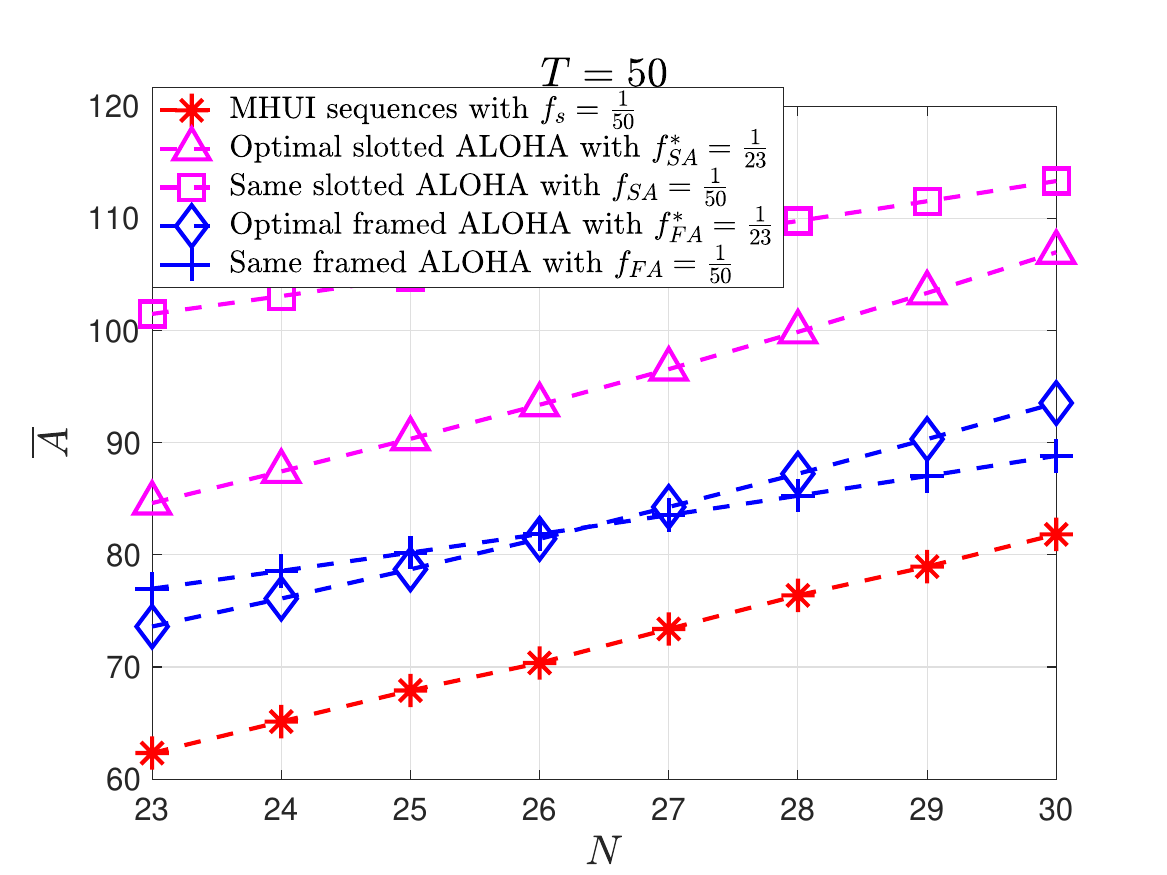}
\caption{AoI performance under various schemes with $T=50$ and $N$ increasing from 23 to 30.}
\label{Fig: check scalability}
\end{figure}

We also conduct simulations to  evaluate the scalability of the proposed scheme. We use one case in Fig.~\ref{Fig: Simulation T=50} for instance, $N=23$ and $T=50$. In this case, both the sequence scheme and the two ALOHA schemes are optimized for the given parameters $N=23$ and $T=50$. The two ALOHA schemes with duty factor equivalent to that of the sequence scheme are also considered. 
Subsequently, we increase 
$N$ from 23 to 30, maintaining the schemes as they are, to observe how the AoI performance is influenced. Notably, for the sequence scheme, since there are only 23 unique sequences available, when 
$N$ exceeds 23, some users must utilize the same sequence. We let the newly added users randomly choose their sequences from the 23  sequences.
As illustrated in Fig.~\ref{Fig: check scalability}, the AoI performance under all schemes deteriorates as the number of newly added users increases.
For the sequence scheme, when 
$N$ reaches 30, the $\overline{A}$ increases to 81.83, marking a 23.8\% increase from the $\overline{A}$ of 62.33 with 
$N=23$. Nevertheless, the sequence scheme continues to outperform all other schemes. For instance, at 
$N=25$, the  $\overline{A}$ for the sequence scheme is 67.9, which is 15.9\% lower than that under the optimal framed ALOHA scheme, which is 78.7. Consequently, the sequence scheme demonstrates the best scalability among the compared schemes.

\vspace{-0.5cm}
\section{Conclusion} \label{sec:conclusion}
In this paper, we employ a sequence scheme to schedule transmissions for periodically generated packets in a system without channel sensing, feedback information from the AP, and time synchronization. Analysis on AoI under the sequence scheme has been conducted, and a mathematical tool based on integer partitions has been developed to improve the 
computational efficiency of average AoI. 
Besides, we obtain low-complexity closed-form expressions on average AoI for two specific scenarios. Based on the analytical results, we propose an algorithm to optimize sequence construction parameters, which enables AoI performance improvement and energy consumption reduction. 
We also perform simulations to compare our proposed sequence scheme with  slotted ALOHA and framed ALOHA. 
Results show that our proposed sequence scheme achieves superior AoI performance and enhanced energy efficiency.

\vspace{-0.3cm}
\appendices 

\section{Proof of Lemma~\ref{lemma:1-position}}
\label{appendix:lemma:1-position}
\begin{IEEEproof}
In case (i), $\beta=TL$,  $\bm{r}_i=[\underbrace{\bm{s}_i\ \ldots \ \bm{s}_i}_{T}]$, and $w'=Tw$.  
Given the characteristic set $\mathcal{I}_i=\{x_0,x_1,\ldots,x_{w-1} \}$ of $\bm{s}_i$, the characteristic set of $\bm{r}_i$ is in the form $\mathcal{I}'_i=\{aL+x_b: a \in \mathbb{Z}_{T}, b\in \mathbb{Z}_w\}$.
Fix $b\in\mathbb{Z}_w$.
Denote by $\sigma'_a$ the 1-position of the ``1'' in $aL+x_k$, for $a\in\mathbb{Z}_T$. 
We aim to show that these $\sigma'_a$'s are mutually distinct, and thus $\{\sigma'_0,\sigma'_1,\ldots,\sigma'_{T-1}\}=\{0,1,\ldots,T-1\}$.
Suppose to the contrary that $\sigma'_{a_1}=\sigma'_{a_2}$ for some $a_1<a_2$.
It follows that $a_2L+x_k\equiv a_1L+x_k \bmod T$, which implies that $(a_2-a_1)L\equiv 0 \bmod T$.
This contradicts to $\gcd(T,L)=1$ and $1\leq a_2-a_1<T$.
Hence, $\mathcal{D}=\cup_{b=0}^{w-1}\{aL+x_b \bmod T:\,a\in\mathbb{Z}_T\} = \cup_{b=0}^{w-1}\{0,1,\ldots,T-1\} = \{0^w,1^w,\ldots,(T-1)^w\}$.
The proof for case (ii) can be done by a similar argument as in case (i), and thus is omitted~here.

In case (iii), $\beta=L$, $\bm{r}_i=\bm{s}_i$, and
$w'=w$. Given the characteristic set $\mathcal{I}_i=\{x_0,x_1,\ldots,x_{w-1} \}$ of $\bm{s}_i$, we have $\mathcal{I}'_i=\mathcal{I}_i$.
For the case of $\bm{s}_i \in \{\bm{v}_1, \bm{v}_2, \ldots, \bm{v}_{p}\}$, according to the definition of CRT construction, we have
$x_k \mod{q}$ goes through $\mathbb{Z}_{w}$ as $k$ goes through $\mathbb{Z}_{w}$, for   $x_k \in \mathcal{I}_i$. Therefore, each entry $x_k$ in $\mathcal{I}_i$ can be written in the form of $x_k=qa_{k}+j$, where $a_{k}\in \mathbb{Z}_{w}, j\in \mathbb{Z}_{w}$. 
Then, for the 1-position of the ``1'' in $x_k$, we have $x_k\equiv j \mod{T}$ since $T|q$. For the  case of $\bm{s}_i=\bm{v}_{p+1}$, we have $x_k\equiv 0 \mod{T}$ since $x_k=qa_{k}$, where $a_{k}\in \mathbb{Z}_{w}$. 
This completes the proof.
\end{IEEEproof}

\section{Proof of Theorem~\ref{lemma:bound ES-P}}
\label{appendix:lemma:bound ES-P}
\begin{IEEEproof}
Without loss of generality, we fix $\bm{s}_i$ and shift the other $(N-1)$ sequences. The total number of offset vectors for the $(N-1)$ sequences is given by $L^{N-1}$. 
Let $M_r$ denote the number of offset vectors under which  $r$ ``1''s within $\bm{s}_i$ are successful and the other $(w-r)$ ``1''s are collided. 
Then, we have 
$P_r=\frac{M_r}{\binom{w}{r} L^{N-1}}$. We derive $M_r$ based on the counting steps~below.
    \begin{enumerate}
    	\item  We should find a subset of the $(N-1)$ sequences and shift the sequences in the subset to collide the $(w-r)$ ``1''s in $\bm{s}_i$. The size of the subset, denoted by $n$, should be no less than $(w-r)$ since each sequence can only block at most one ``1'' in $\bm{s}_i$. Thus, we have $n\in\{w-r,w-r+1,\ldots,N-1\}$.
    	For each subset size $n$, there are $ \binom{N-1}{n} $ ways to choose $n$ sequences out of the $(N-1)$ sequences. 
    	\item For each chosen subset of size $n$, we should assign these $n$ sequences to collide the $(w-r)$ ``1''s.  This is equivalent to partitioning a set of $n$ balls into $(w-r)$ distinct boxes such that no box is empty. 
    	The number of such partitions is given by $
    	(w-r)!	S(n,w-r)=\sum_{m=0}^{w-r}(-1)^{m}\binom{w-r}{m}(w-r-m)^{n}
    	$, where $S(n,w-r)$ denotes the  \textit{Stirling number of the second kind}~\cite{rennie1969stirling}.
    	\item In each partition, the number of offsets under which a sequence can block a ``1'' in $\bm{s}_i$ is $w$. It follows that the total number of offsets for the chosen $n$ sequences to collide the $(w-r)$ ``1''s in $\bm{s}_i$ in each partition is~$ w^n $.
     For the remaining $(N-1-n)$ sequences, they should have no conflicts with any ``1'' in $\bm{s}_i$. The total number of offsets for these sequences is $ (L-w^2)^{N-1-n} $.
    	\end{enumerate}
    By combining the above counting steps, we have 
$
M_r=\sum_{n=w-r}^{N-1}\binom{N-1}{n}(w-r)!S(n,w-r)w^n (L-w^2)^{N-1-n}.    
$
This completes the proof.
\end{IEEEproof}

\section{Proof of Lemma~\ref{lemma:EX=EY}}
\label{appendix:lemma:EX=EY}
\begin{IEEEproof}
Consider an event $e\in\mathcal{E}_r$. 
For  the slot-level inter-departure time and the frame-level inter-departure time between two adjacent AoI drops, that is, $Y^e_{j}$ and $X^e_{j}$, their difference   is  $Y^e_j-X^e_j=(T-S^e_j)-(T-S^e_{j+1})$. Therefore, we~have 
$Y^e_j=X^e_j+S^e_{j+1}-S^e_j$.
Note that the distribution of $S^e_j$ is periodic across superframes.  We thus have $
 \sum_{j=1}^{\eta} ( S^e_{j+1}-S^e_j ) 
= S^e_{\eta+1}-S^e_1
=  0$, where $\eta$ denotes 
the total number of AoI drops within a superframe. 
It follows that $\mathbb{E}_j[S^e_{j}]=\mathbb{E}_j[S^e_{j+1}]$. Thus,  we have 
$\mathbb{E}_j[Y^e_j]=\mathbb{E}_j[X^e_j]$. 
\end{IEEEproof}

\section{Proof of Lemma~\ref{lemma:partition-preimage}} \label{appendix:parition-preimage}
\begin{IEEEproof}
A \emph{hatted} $\mathsf{sf}$-word is an $\mathsf{sf}$-word with exactly one symbol ``$\mathsf{s}$'' having a hat.
So, any $e\in\mathcal{E}_r$ can produce $r$ distinct hatted $\mathsf{sf}$-words.
For example, $\mathsf{\hat{s}fsff}$ and $\mathsf{sf\hat{s}ff}$ are obtained from $\mathsf{sfsff}$ by putting one hat on an ``$\mathsf{s}$''.
Let $\widehat{\mathcal{E}}_r$ be the collection of hatted $\mathsf{sf}$-words of length $w$ having $r $ ``$\mathsf{s}$''s. 
Obviously, $|\widehat{\mathcal{E}}_r|=r|\mathcal{E}_r|$.
Define a mapping $\widehat{\theta}:\widehat{\mathcal{E}}_r\to\mathcal{P}^w_r$ as an extension of $\theta$ by just ignoring the hat.
Now, let us count $|\widehat{\theta}^{-1}(\bm{c})|$.
Obviously, $|\widehat{\theta}^{-1}(\bm{c})|=r|\theta^{-1}(\bm{c})|$.
On the other hand, there are $\frac{r!}{c_1!c_2!\cdots c_w!}$ permutations of $r$ integers with $r_1$ ``$1$''s, $r_2$ ``$2$''s, $\ldots$, $r_w$ ``$w$''s.
By replacing each integer $j$ by a word $\mathsf{s}\underbrace{\mathsf{ff}\cdots \mathsf{f}}_{j-1}$, each aforementioned permutation turns out to be an $\mathsf{sf}$-word of length $w$ having exactly $r$ ``$\mathsf{s}$''s, called a canonical $\mathsf{sf}$-word.
Notice that all canonical $\mathsf{sf}$-words are leading by an ``$\mathsf{s}$'', and are mutually distinct.
For each canonical $\mathsf{sf}$-word, we put a hat on the first ``$\mathsf{s}$'' and then cyclically shift it one by one to obtain $w$ different shifted versions, each of which is a hatted $\mathsf{sf}$-word.
So far, we obtain $\frac{w\cdot r!}{c_1!c_2!\cdots c_w!}$ hatted $\mathsf{sf}$-words.
It suffices to show that these hatted $\mathsf{sf}$-words form the set $\widehat{\theta}^{-1}(\bm{c})$, i.e., $|\widehat{\theta}^{-1}(\bm{c})|=\frac{w\cdot r!}{c_1!c_2!\cdots c_w!}$.
This is true because, these hatted $\mathsf{sf}$-words are mutually distinct by definition, and each $\hat{e}\in\widehat{\theta}^{-1}(\bm{c})$ can be reduced to a canonical $\mathsf{sf}$-word by shifting the hatted $\mathsf{s}$ to the first position and removing the hat.
This completes the proof.
\end{IEEEproof}

\section{Proof of Lemma~\ref{lemma:bound ES-e3}} \label{appendix:lemma:bound ES-e3} 
\begin{IEEEproof}
For a partition $\bm{c}=(c_1,c_2,\ldots,c_w)\in \mathcal{P}^w_r$, there are $|\theta^{-1}(\bm{c})|$ events corresponding to it. 
We denote these events by $\epsilon_{1}, \epsilon_{2}, \ldots, \epsilon_{|\theta^{-1}(\bm{c})|}$. 
The distance between the $k$-th and $(k+1)$-th successful ``1''s within $\bm{s}_i$ under event $\epsilon_{u}$ is denoted by $d_k(\epsilon_{u})$, $k\in \{0,1,\ldots,r-1\}$, $u\in \{1,2,\ldots, |\theta^{-1}(\bm{c})|\}$. 
Then,  we have
$
\sum_{e\in\mathcal{E}_r} \sum_{k=0}^{r-1} F(d^e_k)  = \sum_{\bm{c} \in \mathcal{P}^w_r}\sum_{u=1}^{|\theta^{-1}(\bm{c})|} \sum_{k=0}^{r-1} F(d_k(\epsilon_{u}))$.
Consider an event $\epsilon_{u}$.  
Among the distances $d_0(\epsilon_{u}),d_1(\epsilon_{u}),\ldots, d_{r-1}(\epsilon_{u})$, there are $c_{j}$ entries each of which is consisting of $j$ consecutive entries in $\{ \ell_0,\ell_1,\ldots,\ell_{w-1} \}$, $j\in \{1,2,\ldots,w \} $. They can be expressed as
$\{d_{c_{j},m}(\epsilon_{u}): d_{c_{j},m}(\epsilon_{u})=\ell_{p_h}+\ell_{p_h\oplus_w 1}  + \cdots + \ell_{p_h\oplus_w j-1}, m\in \{0,1,\ldots,c_{j}-1 \} , h \in \{0,1,\ldots,r-1 \}\},$
where $p_0,p_1,\ldots,p_{r-1}$ denote the positions of the 
$r$ ``s''s in~$\epsilon_{u}$.

Now we consider the $w$ cyclically shifted versions of $\epsilon_{u}$, which are denoted by $\epsilon^0_{u}, \epsilon^1_{u},\ldots,
\epsilon^{w-1}_{u}$. Each of them is also corresponding to the partition $\bm{c}$. In $\epsilon^{\delta}_{u}$, the $r$ ``s''s appear in the positions of $p_{0\oplus_w \delta},p_{1\oplus_w \delta},\ldots,p_{r-1\oplus_w \delta}$, $\delta \in \{0,1,\ldots,w-1\}$.  It follows that in $\epsilon^{\delta}_{u}$, the $c_{j}$ entries that are consisting of $j$ consecutive entries in $\{ \ell_0,\ell_1,\ldots,\ell_{w-1} \}$, can be expressed as 
$\{d_{c_{j},m}(\epsilon^{\delta}_{u}): d_{c_{j},m}(\epsilon^{\delta}_{u})=\ell_{p_h\oplus_w \delta}+\ell_{p_h\oplus_w 1+\delta}  + \cdots + \ell_{p_h\oplus_w j-1+\delta}, m\in \{0,1,\ldots,c_{j}-1 \}, h\in \{0,1,\ldots,r-1 \} \}.$
As $ \sum_{\delta=0}^{w-1}  F(\ell_{p_h\oplus_w \delta}+\ell_{p_h\oplus_w 1+\delta}  + \cdots + \ell_{p_h\oplus_w j-1+\delta}) =\sum_{k=0}^{w-1} F(\ell_k + \ell_{k\oplus_w 1}+\cdots+\ell_{k\oplus_w j-1})=b_j$ for any $h\in \{0,1,\ldots,r-1 \}$, 
 we have 
$
  \sum_{m=0}^{c_{j}-1}   \sum_{\delta=0}^{w-1} F(d_{c_{j},m}(\epsilon^{\delta}_{u})) =c_{j} b_j.
$
Thus, $ \sum_{e\in\mathcal{E}_r} \sum_{k=0}^{r-1} F(d^e_k)$ can be calculated~as 
\begin{align*}
\sum_{e\in\mathcal{E}_r} \sum_{k=0}^{r-1} F(d^e_k)  &  = \sum_{\bm{c} \in \mathcal{P}^w_r}  \frac{|\theta^{-1}(\bm{c})|}{w}\sum_{j=1}^w  \sum_{\delta=0}^{w-1} \sum_{m=0}^{c_{j}-1} F(d_{c_{j},m}(\epsilon^{\delta}_{u})) \\
&=\sum_{\bm{c} \in \mathcal{P}^w_r} \frac{|\theta^{-1}(\bm{c})|}{w} \sum_{j=1}^w c_j b_j  ,
\end{align*}
where $b_j$ is given by \eqref{eq:b_k-e1}.
This completes the proof.
\end{IEEEproof}

\section{Proof of Lemma~\ref{lemma:c_q}} \label{appendix:c_q} 
\begin{IEEEproof}
We denote the coefficient of $x^w y^r$ in $\Psi(x,y)$ by $C_{\Psi}(w,r)$. We have $ C_{\Psi}(w,r)  =\sum_{\bm{c} \in \mathcal{P}^w_r}\frac{1}{c_{1}!c_{2}!\cdots c_{w}!}$,
where $\bm{c}=(c_{1},c_{2},\ldots,c_{w})$, $\sum_{j=1}^w c_{j}=r$,  $\sum_{j=1}^w jc_{j}=w$. 
According to the definition of $Q(x,y)$, 
$C_Q(w,r)$ is given by 
\begin{align} \label{eq:ES-bound-eG1}
    C_Q(w,r) = \sum_{j=1}^{w} b_j C_{\Psi}(w-j, r-1).
\end{align}
Therefore, we have $   C_Q(w,r)=\sum_{j=1}^{w} b_j \sum_{\bm{c} \in \mathcal{P}^{w-j}_{r-1}}\frac{1}{c_{1}!c_{2}!\cdots c_{w-j}!}$.
By induction on $\mathcal{P}_{r}^w$, we have $\mathcal{P}_{r}^w= \cup_{j\in \{1,2,\ldots,w\}} \{ (c_{1},\ldots,c_{j} +1, \ldots, c_{w}):  (c_{1},\ldots,c_{j} , \ldots,  c_{w-j}) \in\mathcal{P}^{w-j}_{r-1} \}$. Thus,
\begin{align}
  C_Q(w,r)  & =\sum_{j=1}^{w}  \sum_{\bm{c} \in \mathcal{P}^{w-j}_{r-1}}  \frac{(c_{j} +1)b_j}{c_{1}!\cdots (c_{j} +1)! \cdots c_{w-j}!}  \\ &=\sum_{\bm{c} \in \mathcal{P}^w_r} \frac{\sum_{j=1}^w c_{j} b_j}{c_{1}!c_{2}!\cdots c_{w}!}. \label{eq:lemma6-e2}
\end{align}
This completes the proof.
\end{IEEEproof}

\section{Proof of Lemma~\ref{lemma:c_q2}} \label{appendix:c_q2} 
\begin{IEEEproof}
For $r=1$, we have $C_Q(w,r)=b_w$ by simply plugging $\bm{c}=(c_1,c_2,\ldots,c_w)=(0,0,\ldots,1)$ into \eqref{eq:lemma6-e2}. For $r\in \{2,\ldots,w\}$, we will calculate $C_Q(w,r)$ based on~\eqref{eq:ES-bound-eG1}. 
We consider $C_{\Psi}(w-j,r-1)$ at first. For $\Psi(x,y)$,
 we have 
$       \Psi(x,y) =\text{e}^{y \sum_{i=1}^{\infty}x^i} 
        =\text{e}^{\frac{xy}{1-x}} 
        =1+\frac{\frac{xy}{1-x}}{1!} +\frac{(\frac{xy}{1-x})^2}{2!}+\cdots 
        =1+yx(1+x+x^2+\cdots)+\frac{1}{2!}y^2 x^2 (1+x+x^2+\cdots)^2 +\cdots.
  $
Therefore, $C_{\Psi}(w-j,r-1)$ equals $\frac{1}{(r-1)!}$ multiplying the coefficient of $x^{w-j-(r-1)} $ in $(1+x+x^2+\cdots)^{r-1}$, which is denoted by 
 $C_x(w-j-(r-1))$. For $C_x(w-j-(r-1))$, we consider the generating function $\frac{1}{1-x}=1+x+x^2+\cdots$. After taking the $(r-2)$-th derivative, we have 
   $\frac{(r-2)!}{(1-x)^{r-1}} 
    =(r-1)!+\frac{r!}{1!}x+\cdots+\frac{(w-j-1)!}{(w-j-(r-1))!}x^{w-j-(r-1)}+\cdots $.
    It follows that $C_x(w-j-(r-1))=\frac{(w-j-1)!}{(r-2)!(w-j-(r-1))!}$. 
    Therefore,  
    \begin{align*}
        C_{\Psi}(w-j,r-1)=\frac{(w-j-1)!}{(r-1)!(r-2)!(w-j-(r-1))!},
    \end{align*}
    for $j \in \{1,2,\ldots,w \}$.
    This completes the proof.
\end{IEEEproof}

\section{Proof of Lemma~\ref{lemma:bound ES-e1}} \label{appendix:lemma:bound ES-e1} 
\begin{IEEEproof}
In this case, $\beta=TL$,  $\bm{r}_i=[\underbrace{\bm{s}_i\ \ldots \ \bm{s}_i}_{T}]$, and $w'=Tw$. 
Consider an event $e \in \mathcal{E}_r$ where the positions of the $r$ successful ``1''s within $\bm{s}_i$ are denoted by $\{ x^s_k: k \in \mathbb{Z}_r\} \in \mathcal{I}_i$.
Then there are $r'=rT$ successful ``1''s within $\bm{r}_i$ in total, 
whose positions are given by $\{aL+x^s_k: a \in \mathbb{Z}_{T}, k\in \mathbb{Z}_r \} \in \mathcal{I}'_i$. 
For the successful ``1'' in $aL+x^s_k$, $k\in \mathbb{Z}_r$, we denote the end of the frame it belongs to by $\varepsilon^e_{a,k} $. 
When $k=r-1$, we have $aL+x^s_{k+1}=(a+1)L+x^s_{0}$, $\varepsilon^e_{a,k+1} =\varepsilon^e_{a+1,0} $. 
The distance between the ``1''s in $aL+x^s_k$ and $aL+x^s_{k+1}$ is given by $d^e_k$.
We denote the inter-departure time between $\varepsilon^e_{a,k} $ and $\varepsilon^e_{a,k+1} $ by $\chi^e_{a,k}$. 
Given $d^e_k$, $\chi^e_{a,k}$ can only take two values, $\left \lceil \frac{d^e_k}{T} \right\rceil  T$ and $ \left  \lfloor \frac{d^e_k}{T} \right \rfloor T$, depending on the 1-position of the ``1'' in $aL+x^s_{k+1}$, which is denoted by $\sigma_{a,k+1}$. 
We illustrate the relationship between $\chi^e_{a,k}$ and $\sigma_{a,k+1}$ in Fig.~\ref{Fig.illustration of X and sigma}.
The condition for $\chi^e_{a,k}=\left \lceil \frac{d^e_k}{T} \right\rceil  T$ is $d^e_k \geq \left(\sigma_{a,k+1} +1 +\left  \lfloor \frac{d^e_k}{T} \right \rfloor T\right)$. That is, 
 \begin{equation} \label{eq:distribution of xk}
     \chi^e_{a,k}=\begin{cases}
     \left \lceil \frac{d^e_k}{T} \right\rceil  T, \text{ if } \sigma_{a,k+1} \in \left[0, d^e_k - \left\lfloor \frac{d^e_k}{T} \right\rfloor T -1 \right], \\
       \left  \lfloor \frac{d^e_k}{T} \right \rfloor T, \text{ if } \sigma_{a,k+1}  \in \left[ d^e_k -\left \lfloor \frac{d^e_k}{T} \right \rfloor T   , T-1\right]. \\
     \end{cases}
 \end{equation}

\begin{figure*}[h] 
	\centering 
\includegraphics[width=\textwidth]{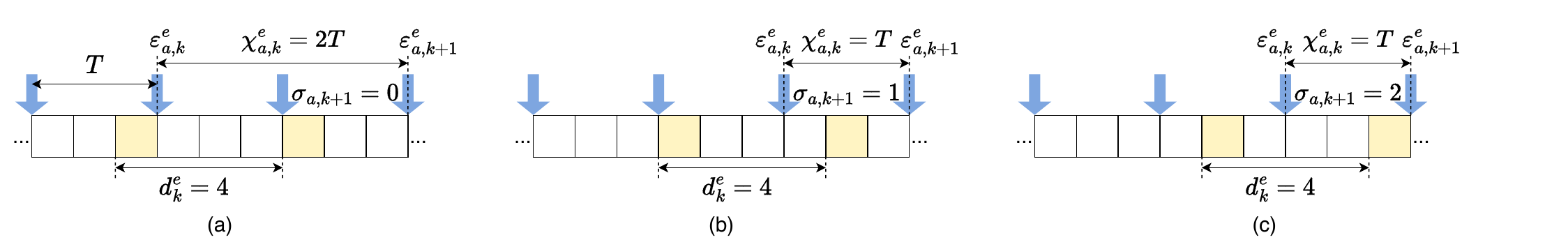} 
	\caption{Illustration of $ \chi^e_{a,k}$ and $\sigma_{a,k+1}$.
 In the given scenario, $T=3$, $d^e_k=4$.  The two successful ``1''s in $aL+x^s_k$ and $aL+x^s_{k+1}$ are marked in yellow. As observed, $\chi^e_{a,k}=2T$  when $\sigma_{a,k+1}=0$, and $\chi^e_{a,k}=T$  when $\sigma_{a,k+1}=1$ or $2$.
 }
	\label{Fig.illustration of X and sigma}  

\end{figure*}

Let $\eta$ denote the total number of  AoI drops within $\bm{r}_i$. Then, 
\begin{equation} \label{eq:EX-coprime}
  \mathbb{E}_j[X^e_j]= \frac{ \sum \limits_{k=0}\limits^{r-1}\sum\limits_{a=0}\limits^{T-1} \chi^e_{a,k}}{\eta},  
\mathbb{E}_j[(X^e_j)^2]=\frac{ \sum\limits_{k=0}\limits^{r-1} \sum\limits_{a=0}\limits^{T-1} \left(\chi^e_{a,k}\right)^2}{\eta} .   
\end{equation}

Let $\mathcal{S}^e_{a,k+1}$ be the service time of the ``1'' in $aL+x^s_{k+1}$. If this ``1'' is the first successful ``1'' within the frame it belongs to, then  $\mathcal{S}^e_{a,k+1}=\sigma_{a,k+1}$. Otherwise,  $\chi^e_{a,k}=0$. Therefore,  
\begin{equation} \label{eq:ESX-coprime}
\mathbb{E}_j[S^e_{j+1} X^e_j]=\frac{\sum\limits_{k=0}\limits^{r-1}\sum\limits_{a=0}\limits^{T-1}\mathcal{S}^e_{a,k+1}\chi^e_{a,k}}{\eta}=\frac{\sum\limits_{k=0}\limits^{r-1}\sum\limits_{a=0}\limits^{T-1}\sigma_{a,k+1}\chi^e_{a,k}}{\eta}.  
\end{equation}

According to Lemma~\ref{lemma:1-position}, given $k$, the 1-positions of the $T$ ``1''s in $\{aL+x^s_k: a\in \mathbb{Z}_{T} \}$ are distinct and go through $\{0,1,\ldots,T-1 \}$. Then,
according to \eqref{eq:distribution of xk}, given $k$,
  among all the $T$    $\chi^e_{a,k}$'s for $a \in \mathbb{Z}_{T}$, $ \left( T-f^e_k-1 \right)$  of them are of length $\left\lfloor \frac{d^e_k}{T} \right\rfloor T$, and the others are of length $\left\lceil \frac{d^e_k}{T} \right \rceil T$. Thus, \eqref{eq:EX-coprime-X-e1}, \eqref{eq:EX-coprime-X-e2} and~\eqref{eq:SX-coprime-e3} can be obtained. The proof is completed by combining \eqref{eq:EX-coprime}, \eqref{eq:ESX-coprime}, \eqref{eq:EX-coprime-X-e1}, \eqref{eq:EX-coprime-X-e2} and~\eqref{eq:SX-coprime-e3}.
\end{IEEEproof}

\begin{figure*}
\begin{gather}
\sum_{k=0}^{r-1}\sum_{a=0}^{T-1} \chi^e_{a,k} 
= \sum_{k=0}^{r-1} \left( \left \lfloor \frac{d^e_k}{T} \right \rfloor T \left( T-f^e_k-1 \right)  +  \left \lceil \frac{d^e_k}{T} \right \rceil T \left( f^e_k+1 \right) \right) = LT, \label{eq:EX-coprime-X-e1} \\
  \sum_{k=0}^{r-1} \sum_{a=0}^{T-1} \left(\chi^e_{a,k}\right)^2 =  \sum_{k=0}^{r-1}  \left( \left( \left \lfloor \frac{d^e_k}{T} \right \rfloor T \right) ^2 \left( T-f^e_k-1  \right) + \left( \left \lceil \frac{d^e_k}{T} \right \rceil T \right)^2 \left( f^e_k+1 \right) \right) = \sum_{k=0}^{r-1} T^2 \left( \left \lfloor \frac{d^e_k}{ T} \right \rfloor (2d^e_k -T) - \left \lfloor \frac{d^e_k}{ T} \right \rfloor^2 T +d^e_k \right), \label{eq:EX-coprime-X-e2} 
  \end{gather}
  \vspace{-0.7cm}
  \begin{align}
 \sum_{k=0}^{r-1} \sum_{a=0}^{T-1} \sigma_{a,k+1} \chi^e_{a,k}  &=  \sum_{k=0}^{r-1} \left(\sum_{\sigma_{a,k+1}=0}^{f^e_k} \left \lceil \frac{d^e_k}{T} \right \rceil T \sigma_{a,k+1} + \sum_{\sigma_{a,k+1}=f^e_k+1}^{T-1} \left \lfloor \frac{d^e_k}{T} \right \rfloor T \sigma_{a,k+1} \right) \notag \\
 &= \sum_{k=0}^{r-1} \left(\left \lceil \frac{d^e_k}{T} \right \rceil T \left( \frac{f^e_k(f^e_k+1)}{2} \right)+ \left \lfloor \frac{d^e_k}{T} \right \rfloor T \left( \frac{(f^e_k+T)(T-f^e_k-1)}{2} \right) \right). \label{eq:SX-coprime-e3}
 \end{align}
\end{figure*}

\section{Proof of Lemma~\ref{lemma:case2-1}} \label{appendix:lemma:case2-1}
\begin{IEEEproof}
  According to \cite{shum2009design}, the difference sets of UI sequences are disjoint. The difference set of $\bm{v}_1$ is given by $\{\pm 1, \pm 2, \ldots, \pm (p-1) \}$. Therefore, in each of the other sequences $\{\bm{v}_2,\bm{v}_3,\ldots,\bm{v}_{p+1} \}$, there are no two consecutive ``1''s that have a distance less than $(p-1)$. In other words, there is at most one ``1'' in any duration of $p$ slots. This completes the proof.
\end{IEEEproof} 

\section{Proof of Lemma~\ref{lemma:case2-2}} \label{appendix:lemma:case2-2}
\begin{IEEEproof}
According to the definition of CRT construction, under the CRT correspondence, for $g\in \{2,\ldots,p\}$, the characteristic set of sequence $\bm{v}_g$ is equal to 
   $
         \mathcal{I}_g  =\{ u(g,1): u\in \mathbb{Z}_{w}\} 
          = \{ x_u +(u,u): u\in \mathbb{Z}_{w}\},
    $
where $x_u= u(g-1,0)$. 
To show that there is at most one ``1'' in any duration of $q$ slots within $\bm{v}_g$, it suffices to claim that $x_u \neq x_v$, for $u \neq v$, $u, v \in \mathbb{Z}_{w}$.
Suppose to the contrary that $x_u = x_v$, which means
$
    u(g-1,0)=v(g-1,0).$
It follows  that 
$
    (u-v)(g-1)=0 \mod p. 
$
However, this cannot hold due to the fact that $u-v \in   \{\pm 1, \pm 2, \ldots, \pm (w-1) \}$, $w=p$ and $g \neq 1$. 
For $g=p+1$, the characteristic set of sequence $\bm{v}_g$ is equal to  $ \mathcal{I}_g  =\{ (u,0 ): u\in \mathbb{Z}_{w}\}$. Thus,  there is also at most one ``1'' in any duration of $q$ slots within $\bm{v}_{p+1}$.
\end{IEEEproof}

\section{Proof of Lemma~\ref{lemma:A-case2-1}} \label{appendix:lemma:special case 2} 
\begin{IEEEproof}
Consider an event $e\in \mathcal{E}_r$. Note that in the case we consider, each successful ``1'' corresponds to an AoI drop. That is, $Y^e_j =d^e_j$. 
Then,  we have  
$
  \mathbb{E}_j[Y^e_j] = \frac{\sum_{j=0}^{r-1}d^e_j}{r} =\frac{L}{r}, ~
  \mathbb{E}_j[(Y^e_j)^2]  =   \frac{\sum_{j=0}^{r-1}(d^e_j)^2}{r} . 
  $
Further, we have 
$
    \frac{\mathbb{E}_j[(Y^e_j)^2]}{2\mathbb{E}_j[Y^e_j]} = \frac{\sum_{j=0}^{r-1}(d^e_j)^2}{2L}.
$
The total number of AoI drops within a superframe $\bm{r}_i$, $\eta$, is given by $\eta=r\beta/L$.
We have
\begin{equation}
\label{eq:ESY-EY-case2}
\sum_{e\in\mathcal{E}_r}\frac{\mathbb{E}_j[S^e_jY^e_j]}{\mathbb{E}_j[Y^e_j]} = \sum_{e\in\mathcal{E}_r} \frac{1}{\beta} \sum_{j=0}^{\eta-1} S^e_j Y^e_j .
\end{equation}

\vspace{-0.3cm}
For a partition $\bm{c}=(c_1,c_2,\ldots,c_w)\in \mathcal{P}^w_r$, there are $|\theta^{-1}(\bm{c})|$ events corresponding to it. 
These events  are denoted by $\epsilon_{1}, \epsilon_{2}, \ldots, \epsilon_{|\theta^{-1}(\bm{c})|}$. 
Then we have 
\begin{equation} 
\sum_{e\in\mathcal{E}_r} \frac{1}{\beta} \sum_{j=0}^{\eta-1} S^e_j Y^e_j = \sum_{\bm{c} \in \mathcal{P}^w_r}\sum_{u=1}^{|\theta^{-1}(\bm{c})|} \frac{1}{\beta}\sum_{j=0}^{\eta-1} S^j(\epsilon_{u})Y^j(\epsilon_{u}),
\end{equation}
where $S^j(\epsilon_{u})$ denotes the service time for the  $j$-th AoI drop in event $\epsilon_{u}$, and $Y^j(\epsilon_{u})$ denotes the inter-departure time between the $j$-th and $(j+1)$-th AoI drops in event $\epsilon_{u}$, $u\in \{ 1,2,\ldots,|\theta^{-1}(\bm{c})|\} $. 
For the event $\epsilon_{u}$,
we at first consider $S^j(\epsilon_{u})Y^j(\epsilon_{u}) $ within each sequence period $L$. 
Among the $Y^j(\epsilon_{u})$'s within the $v$-th sequence period,
$v\in \{0,1,\ldots,\beta/L-1\}$, there are $c_j$ entries each of which is consisting of $j$ consecutive entries in $\{ \ell_{vw},\ell_{vw+1},\ldots,\ell_{vw+w-1} \}$, $j\in \{1,2,\ldots,w \} $.
 The $c_j$ entries can be expressed as
$\{Y^{j}_{v,m}(\epsilon_{u}): Y_{v,m}^{j}(\epsilon_{u})=\ell_{vw\oplus_{w'} p_m}+\ell_{vw \oplus_{w'}+p_m +1}  + \cdots + \ell_{vw\oplus_{w'} +p_m+j-1}, m\in \{0,1,\ldots,c_j-1 \} \},$
where $p_0,p_1,\ldots,p_{r-1}$ denote the positions of the 
$r$ ``s''s in $\epsilon_{u}$. We use $S_{v,m}^{j}(\epsilon_{u})$ to denote the service time of the AoI drop corresponding to $Y_{v,m}^{j}(\epsilon_{u})$, which is given by $S_{v,m}^{j}(\epsilon_{u})=\sigma_{vw\oplus_{w'} p_m}$. For event $\epsilon_{u}$, we have 
\begin{equation*}
    \frac{1}{\beta}\sum_{j=0}^{\eta-1} S^j(\epsilon_{u})Y^j(\epsilon_{u})=   \frac{1}{\beta}\sum_{v=0}^{\beta/L-1} \sum_{j=1}^w \sum_{m=0}^{c_j-1} S_{v,m}^{j}(\epsilon_{u})Y_{v,m}^{j}(\epsilon_{u}). 
\end{equation*}
Now we consider the $w$ cyclically shifted versions of $\epsilon_{u}$, which are denoted by $\epsilon^0_{u}, \epsilon^1_{u},\ldots,
\epsilon^{w-1}_{u}$. Each of them is also corresponding to the partition $\bm{c}$. In $\epsilon^{\delta}_{u}$, $\delta \in \{0,1,\ldots,w-1 \}$, the $r$ ``s''s appear in the positions of $p_{0\oplus_w \delta},p_{1\oplus_w \delta},\ldots,p_{r-1\oplus_w \delta}$.  It follows that for event $\epsilon^{\delta}_{u}$, among the $Y^j(\epsilon^{\delta}_{u})$'s within the $v$-th sequence period, 
the $c_j$ entries that are consisting of $j$ consecutive entries in $\{ \ell_{vw},\ell_{vw+1},\ldots,\ell_{vw+w-1} \}$, $j\in \{1,2,\ldots,w \} $, can be expressed as 
$\{Y_{v,m}^j(\epsilon^{\delta}_{u}): Y_{v,m}^j(\epsilon^{\delta}_{u})=\ell_{vw \oplus_{w'} p_m+ \delta}+\ell_{vw \oplus_{w'} p_m+1+ \delta }  + \cdots + \ell_{vw \oplus_{w'} p_m +j-1+\delta}, m\in \{0,1,\ldots,c_j-1 \} \}.$ 
For $S^j_{v,m}(\epsilon^{\delta}_{u})$ which  corresponds to $Y^j_{v,m}(\epsilon^{\delta}_{u})$, it is given by $S_{v,m}^j(\epsilon^{\delta}_{u})=\sigma_{vw \oplus_{w'} p_m+ \delta}$.
Thus, 
\begin{align*}
& \sum_{j=1}^w \sum_{m=0}^{c_j-1}  \sum_{v=0}^{\beta/L-1} \sum_{\delta=0}^{w-1}   S^j_{v,m}(\epsilon^{\delta}_{u})Y^j_{v,m}(\epsilon^{\delta}_{u}) \\
= & \sum_{j=1}^w \sum_{m=0}^{c_j-1}  \sum_{v=0}^{\beta/L-1} \sum_{\delta=0}^{w-1}\sigma_{vw \oplus_{w'} p_m+ \delta} \sum_{z=0}^{j-1}\ell_{vw \oplus_{w'} p_m+ \delta +z} \\ 
=&\sum_{j=1}^w c_j   \sum_{k=0}^{w'-1} (\ell_k + \ell_{k\oplus_{w'} 1}+\cdots+\ell_{k\oplus_{w'} j-1}) \sigma_k .
\end{align*}
Then, based on~\eqref{eq:ESY-EY-case2}, we have
\begin{align*}
&\sum_{e\in\mathcal{E}_r}\frac{\mathbb{E}_j[S^e_jY^e_j]}{\mathbb{E}_j[Y^e_j]} \\
= &\sum_{\bm{c} \in \mathcal{P}^w_r}   \frac{|\theta^{-1}(\bm{c})|}{\beta w} \sum_{j=1}^w \sum_{m=0}^{c_j-1}  \sum_{v=0}^{\beta/L-1} \sum_{\delta=0}^{w-1}   S^j_{v,m}(\epsilon^{\delta}_{u})Y^j_{v,m}(\epsilon^{\delta}_{u}) \\
=&\sum_{\bm{c} \in \mathcal{P}^w_r}   \frac{|\theta^{-1}(\bm{c})|}{\beta w}  \sum_{j=1}^w c_j   \sum_{k=0}^{w'-1}\zeta_{w'}(k,j) \sigma_k .
\end{align*} 

This completes the proof.
\end{IEEEproof}

\ifCLASSOPTIONcaptionsoff
  \newpage
\fi

\bibliographystyle{IEEEtran}
\bibliography{ref_new}

\end{document}